\newif\ifisarxive
\isarxivetrue  

\ifisarxive
    \documentclass{article}
    \pdfoutput=1
    \usepackage{arxiv}
    \usepackage{natbib}
    \usepackage[toc,page]{appendix}
\else
    \documentclass[numsec, webpdf, modern, medium, namedate]{oup-authoring-template}
    \onecolumn 
\fi

\usepackage[utf8]{inputenc}

\usepackage{amsmath}       
\usepackage{amsfonts} 
\usepackage{enumerate}
\usepackage{hyperref}
\usepackage{pdflscape} 
\usepackage{comment} 
\usepackage{graphicx}       
\usepackage{subcaption}
\usepackage{xcolor}         
\usepackage{tikz}
\usetikzlibrary{shapes.geometric, arrows, positioning, calc}
\definecolor{blue}{RGB}{0, 0, 213}
\definecolor{myblue}{RGB}{40, 75, 148}
\definecolor{gold3}{RGB}{205, 173, 0}
\usepackage[acronym,section,nonumberlist]{glossaries}
\makenoidxglossaries
\newacronym{dif}{DIF}{Differential Item Functioning}
\newacronym{irt}{IRT}{Item Response Theory}
\newacronym{irc}{IRC}{Item Response Curve}
\newacronym{pl}{PL}{Parameter Logistic}
\newacronym{rmse}{RMSE}{Root Mean Squared Error}
\DeclareMathOperator{\Prob}{\textsf{P}}
\DeclareMathOperator{\E}{\textsf{E}}
\usepackage{stackrel}
\newcommand{\gotodN}{\stackrel[N \goto \infty]{\mathcal{D}}{\longrightarrow}}
\newcommand{\goto}{\rightarrow}
\newcommand{\abs}[1]{\left|{#1}\right|}
\usepackage{tabularx}
\usepackage{booktabs}
\usepackage{siunitx}
\usepackage{multirow}
\usepackage{diagbox}
\DeclareRobustCommand\Rcode{\bgroup\@codex}
\def\@codex#1{\texorpdfstring%
{{\normalfont\ttfamily #1}}%
{#1}\egroup}

\let\pkg=\strong
\newcommand{\CRANpkg}[1]{\href{https://CRAN.R-project.org/package=#1}{\pkg{#1}}}%

\ifisarxive
    \title{A novel nonparametric framework for DIF detection using kernel-smoothed item response curves}
    \renewcommand{\shorttitle}{Nonparametric Comparison of Item Response Curves}
    \author{Ad\'ela Hladk\'a$^{1}$, Patr\'icia Martinkov\'a$^{1,2}$\\
    \small $^{1}$ Institute of Computer Science of the Czech Academy of Sciences, Prague, Czech Republic\\
    \small $^{2}$ Faculty of Education, Charles University, Prague, Czech Republic\\
\date{November 24, 2025}}
\else
    \journaltitle{The Journal of the Royal Statistical Society: Series A (Statistics in Society)}
    \DOI{}
    \copyrightyear{2025}
    \pubyear{2025}
    \access{}
    
    \firstpage{1}
    
    \title[Nonparametric Comparison of Item Response Curves]{A novel nonparametric framework for DIF detection using kernel-smoothed item response curves}
    
    \author[1, $\ast$]{Adéla Hladká\ORCID{0000-0002-9112-1208 }}
    \author[1, 2]{Patrícia Martinková\ORCID{0000-0003-4754-8543}}
    
    \authormark{Hladká and Martinková}
    
    \address[1]{\orgdiv{Department of Statistical Modelling}, \orgname{Institute of Computer Science of the Czech Academy of Sciences}, \orgaddress{\street{Pod Vodárenskou věží 271/2, Prague}, \postcode{182 07}, \country{Czech Republic}}}
    \address[2]{\orgdiv{Institute for Research and Development of Education}, \orgname{Faculty of Education, Charles University}, \orgaddress{\street{Myslíkova 7, Prague}, \postcode{110 00}, \country{Czech Republic}}}
    
    \corresp[$\ast$]{Corresponding author: \href{email:martinkova@cs.cas.cz}{martinkova@cs.cas.cz}}
\fi

\begin{document}


\ifisarxive
    \maketitle
    \begin{abstract}
        This study introduces a novel nonparametric approach for detecting Differential Item Functioning (DIF) in binary items through direct comparison of Item Response Curves (IRCs). Building on prior work on nonparametric comparison of regression curves, we extend the methodology to accommodate binary response data, which is typical in psychometric applications. The proposed approach includes a new estimator of the asymptotic variance of the test statistic and derives optimal weight functions that maximise local power. Because the asymptotic distribution of the resulting test statistic is unknown, a wild bootstrap procedure is applied for inference. A Monte Carlo simulation study demonstrates that the nonparametric approach effectively controls Type I error and achieves power comparable to the traditional logistic regression method, outperforming it in cases with multiple intersections of the underlying IRCs. The impact of bandwidth and weight specification is explored. Application to a verbal aggression dataset further illustrates the method’s ability to detect subtle DIF patterns missed by parametric models. Overall, the proposed nonparametric framework provides a flexible and powerful alternative for detecting DIF, particularly in complex scenarios where traditional model-based assumptions may not be applicable.
    \end{abstract}
\fi


\section{Introduction}

Multi-item assessments play a significant role in various areas of our everyday lives, intervening in education, psychology, healthcare, and other applied fields \citep{brennan2006educational, haladyna2011handbook, martinkova2023computational}. In the educational context, they are used to assess academic performance, certify student qualifications, measure knowledge and skill proficiency, and administer admission tests, as well as national and international large-scale student exams. In psychology, these assessments help measure intelligence, personality traits, and attitudes
. In health-related fields, they are employed for measuring outcomes such as fatigue, depression, pain, quality of life, and overall well-being. Similar instruments are also further applied in employment selection and promotion processes, peer reviews of academic work, and the evaluation of grant proposals.

While decisions based on such assessments are often guided by total scores, a detailed analysis at the item level provides deeper insight. Item functioning can be described by \gls{irc}, which expresses the probability of a specific response to an item as a function of an individual's latent trait and, potentially, additional covariates. Parametric modelling frameworks, such as \gls{irt} \citep{vanderlinden1997handbook, vanderlinden2018handbook} and score-based regression models \citep{martinkova2023computational}, are widely used to estimate \glspl{irc}. 

However, any parametric approach risks oversimplification when the underlying model omits critical information. In cases where the true model is highly complex or lacks a clear parametric form, nonparametric methods offer a more flexible alternative. Notable examples include the monotone homogeneity model and double-monotonicity model for binary items \citep{mokken1971theory}, as well as the kernel-smoothing approach employing Nadaraya-Watson weights and rank-based ability estimates \citep{ramsay1991kernel}. 

A crucial aspect of item functioning is that it may vary across different respondent groups. This phenomenon, known as \gls{dif}, occurs when respondents from different groups but with the same underlying trait have different probabilities of giving a certain response to an item. Parametric approaches, such as \gls{irt} models and score-based regression methods, can be extended to account for the effect of covariates (e.g., group-membership variable) and applied for \gls{dif} detection \citep{lord1980applications, raju1988area, swaminathan1990detecting, drabinova2017detection, hladka2023combining, hladka2025estimation}. Yet, to our knowledge, a nonparametric \gls{dif} detection directly based on comparing \glspl{irc} has not been systematically explored. 

To address this gap, we introduce a new kernel-smoothing-based framework for the nonparametric comparison of \glspl{irc} across groups. Our method adapts and extends the general approach of \citet{srihera2010nonparametric} to the binary-item setting, developing it specifically for the purpose of detecting \gls{dif}. We further propose several methodological variants designed to accommodate different \gls{dif} types, and we demonstrate their performance in a systematic comparison with the widely used logistic regression method \citep{swaminathan1990detecting}. 

The remainder of this paper is structured as follows. Section \ref{sec:methodology} details the estimation procedure, the general test statistic, and several weight-function strategies, including optimal weights designed to maximise test power, along with their estimates and asymptotic properties of the resulting test statistic. Section \ref{sec:simulation} presents a simulation study evaluating the proposed framework in comparison to the traditional parametric approach based on logistic regression. Section \ref{sec:real_data} illustrates the methodology using an empirical dataset from a questionnaire about verbal aggression. Section \ref{sec:discussion} concludes with a discussion of the findings, practical recommendations, and directions for future work.  


\section{Methodology}\label{sec:methodology}

\gls{dif} and its detection are closely connected to the broader problem of describing the relationship between respondents' responses $\boldsymbol{Y_i} = (Y_{1i}, \dots, Y_{ni})$ to item $i$ and their abilities $\boldsymbol{\theta} = (\theta_1, \dots, \theta_n)$, where $n$ is a number of respondents. This relationship can generally be expressed through a regression function $m_i$ for item $i$:
\begin{align*}
    \boldsymbol{Y_i} = m_i(\boldsymbol{\theta}) + \boldsymbol{\epsilon_i}, \quad \E(\boldsymbol{\epsilon_i}| \boldsymbol{\theta}) = \boldsymbol{0}
\end{align*}
This work focuses on binary outcomes $\boldsymbol{Y_i}$, in which case this relationship can be reformulated as
\begin{align*}
    \E(\boldsymbol{Y_i}|\boldsymbol{\theta}) = \Prob(\boldsymbol{Y_i} = 1|\boldsymbol{\theta}) = m_i(\boldsymbol{\theta}).
\end{align*}
We estimate \glspl{irc}, i.e., the function $m_i(\cdot)$, using the nearest-neighbor kernel-smoothing approach of \citet{srihera2010nonparametric} across two groups: reference ($g = 0$) and focal ($g = 1$). Let $\mathcal{P}_g$ denote the set of respondents in group $g$ of size $n_g$, such that $n_0 + n_1 = n$ is the total sample size. For a given item $i$, let $Y_{pi}$ be the binary response of respondent $p$ and $\theta_{p}$ their ability (e.g., standardised total test score or other matching criterion). Define $\hat{F}_{ig}(x)$ as the empirical distribution function of $\theta_{p}$ in group $g$:
\begin{align*}
    \hat{F}_{g}(x) = \frac{1}{n_g} \sum_{p \in \mathcal{P}_g} \mathbf{1} [\theta_{p} \leq x].
\end{align*}
The nearest neighbor estimate of the \gls{irc} $m_{ig}$ for item $i$ for group $g$ is then given by:
\begin{align}\label{eq:estimate}
    \hat{m}_{ig}(x) = \sum\limits_{p \in \mathcal{P}_g} Y_{pi} W_{pig}(x),
\end{align}
where the weights $W_{pig}(x)$ are defined as: 
\begin{align}\label{eq:weights}
    W_{pig}(x) = \frac{K\left(\frac{\hat{F}_{g}(\theta_{p}) - \hat{F}_{g}(x)}{h}\right)}{\sum\limits_{k \in \mathcal{P}_g} K\left(\frac{\hat{F}_{g}(\theta_{k}) - \hat{F}_{g}(x)}{h}\right)}. 
\end{align}
$K(\cdot)$ is a twice continuously differentiable, symmetric, non-negative kernel function, which is non-decreasing for $u < 0$ with a compact support and $\int K(u) \ \mathrm{d}u = 1$. Examples include the Epanechnikov kernel $K(u) = \frac{3}{4}(1 - u^2), \abs{u} \leq 1$ \citep{epanechnikov1969non}, and the uniform kernel $K(u) = \frac{1}{2}, \abs{u} \leq 1$. The bandwidth parameter $h$ satisfies $nh^3 \rightarrow \infty$ and $nh^4 \rightarrow 0$ as $n \rightarrow \infty$ \cite[see][p. 2042]{srihera2010nonparametric}. Therefore, the parameter $h$ is assumed to take the value of $n^{-\zeta}$, where $\zeta \in \left(\frac{1}{4}, \frac{1}{3}\right)$\label{assumption} and $n$ has the order of $n_0$ and $n_1$.

\begin{figure}[htb]
  \centering
  \begin{subfigure}[t]{.32\textwidth}
    \centering
    \includegraphics[width = \textwidth]{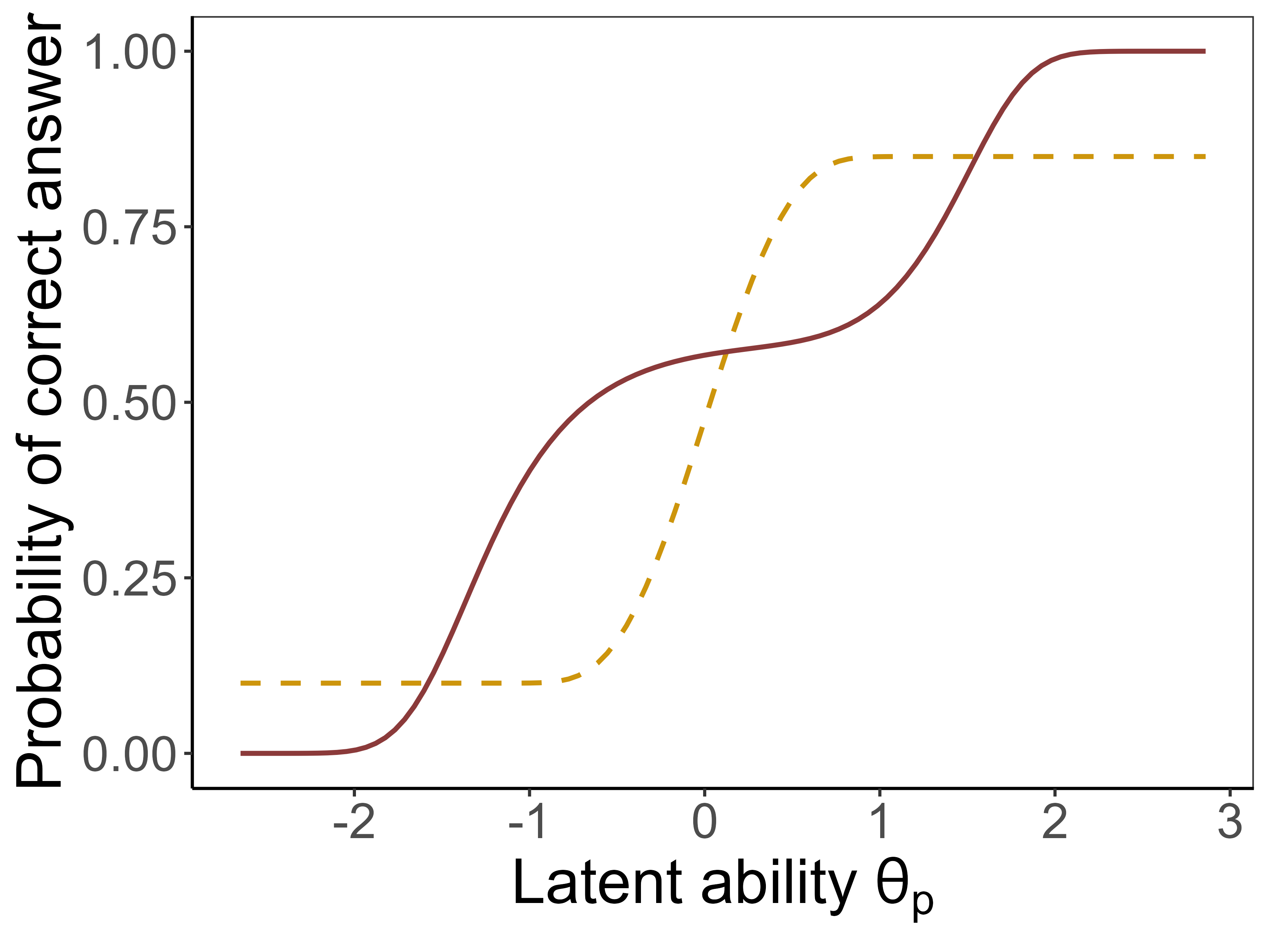}
    \caption{Underlying IRCs}\label{fig:illustration_NN_example:A}
  \end{subfigure}
  \begin{subfigure}[t]{.32\textwidth}
    \centering
    \includegraphics[width = \textwidth]{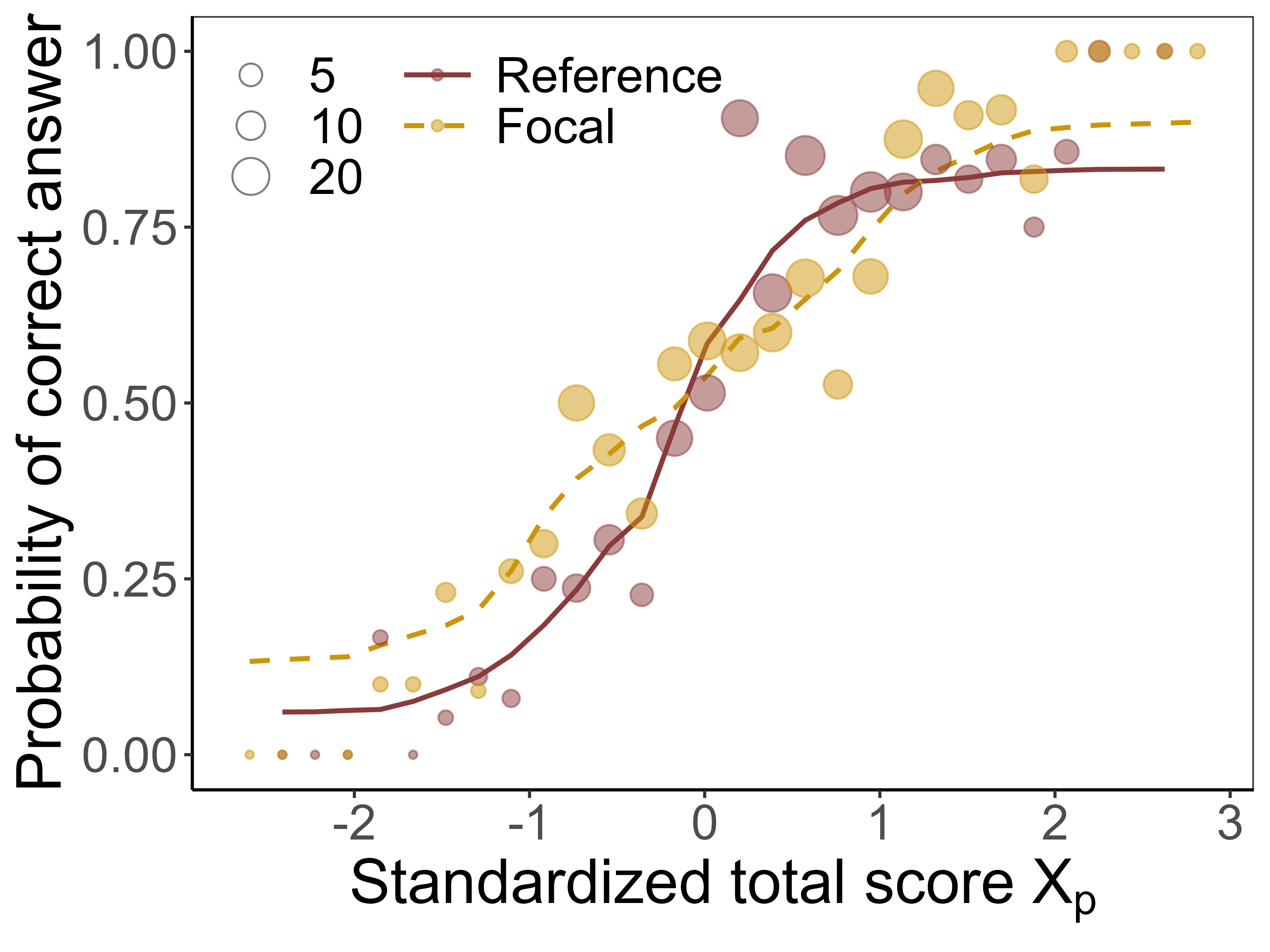}
    \caption{Nearest neighbours}\label{fig:illustration_NN_example:B}
  \end{subfigure}
  \begin{subfigure}[t]{.32\textwidth}
    \centering
    \includegraphics[width = \textwidth]{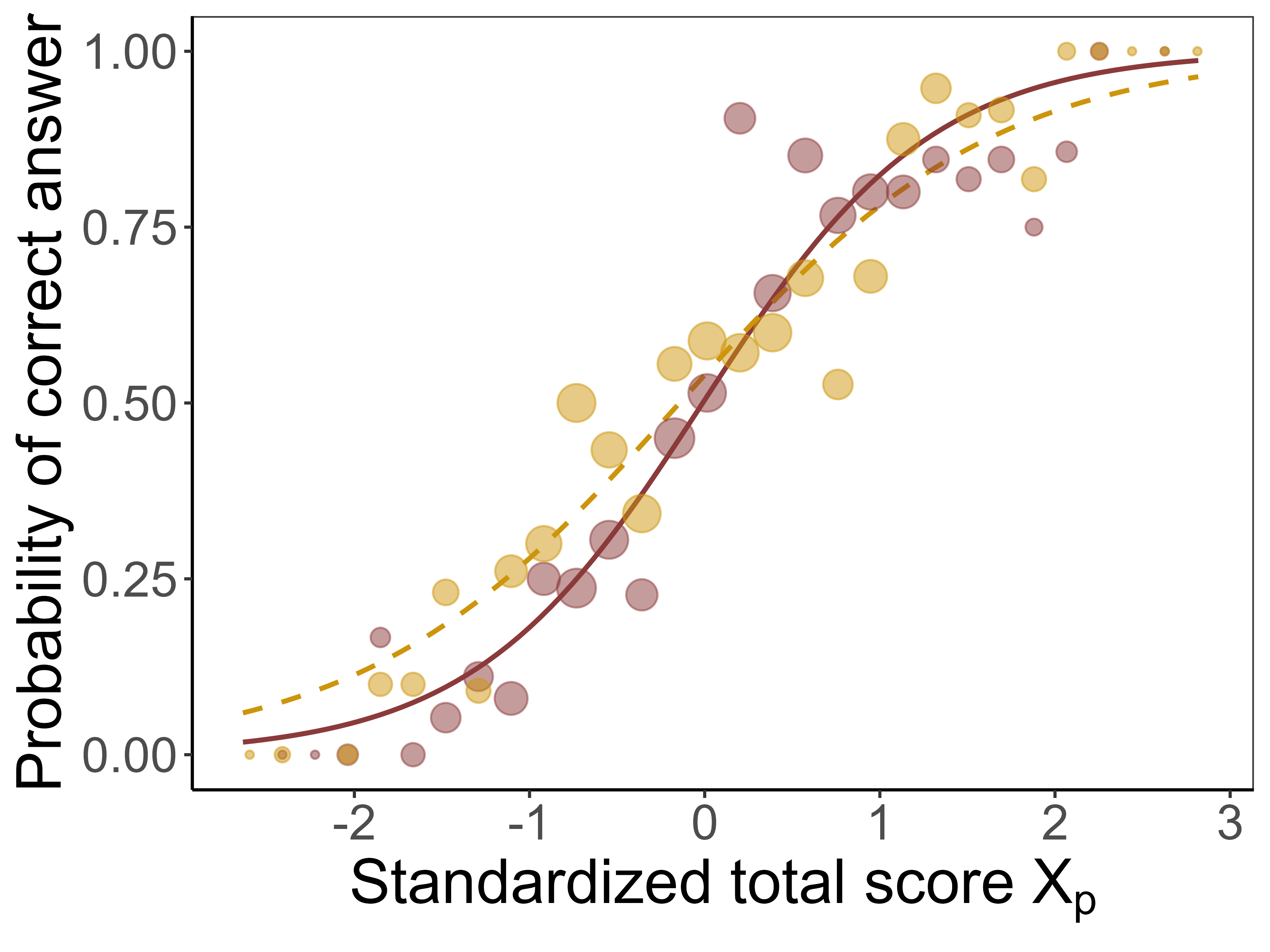}
    \caption{Logistic regression}\label{fig:illustration_NN_example:C}
  \end{subfigure}
  \caption{Example of nearest neighbour and logistic regression estimates of IRCs; curves are accompanied by points representing empirical probabilities. }
\end{figure}

Kernel smoothing is advantageous here because it makes no assumptions about the functional form of $m_{ig}(x)$, making it applicable even when \glspl{irc} have complex shapes. For an illustration, for \glspl{irc} with multiple inflection points (Figure~\ref{fig:illustration_NN_example:A}), the nearest neighbor estimate \eqref{eq:estimate} applied to simulated binary data provides a closer match to the true curves (Figure~\ref{fig:illustration_NN_example:B}) than the logistic regression approach (Figure~\ref{fig:illustration_NN_example:C}).

\subsection{Test statistic}

Differences in ability distributions between groups can make direct curve comparison problematic. Following \citet[][p. 2040]{srihera2010nonparametric}, we define a common support by averaging the values of the matching criterion $\theta_p$ from the two groups:
\begin{align*}
    \bar{X}_{p_0p_1} = \frac{\theta_{p_0} + \theta_{p_1}}{2}, \quad p_0 \in \mathcal{P}_0, p_1 \in \mathcal{P}_1.
\end{align*}
The proposed general test statistic for item $i$ is then expressed as follows:
\begin{align}\label{eq:nonparametric_test}
    \widehat{T}_i = \frac{1}{n_0 n_1}\sum_{p_0 \in \mathcal{P}_0}\sum_{p_1 \in \mathcal{P}_1} W_i\left(\bar{X}_{p_0p_1}\right)\left[\hat{m}_{i0}\left(\bar{X}_{p_0p_1}\right) - \hat{m}_{i1}\left(\bar{X}_{p_0p_1}\right)\right],
\end{align}
where $W_i(\cdot)$ is a twice continuously differentiable weight function for item $i$. $\widehat{T}_i$ is the weighted average difference between the two \glspl{irc} across the common support. Under the null hypothesis $H_0{:}\,\, m_{i0} \equiv m_{i1}$, i.e., no \gls{dif}, $\widehat{T}_i$ should be close to zero. 


\subsubsection{Asymptotic properties}

\paragraph{Asymptotic variance. } The asymptotic variance of the test statistic \eqref{eq:nonparametric_test} under the null hypothesis is given by 
\begin{align}\label{eq:variance}
    \sigma_i^2 = (1 - \lambda) \rho_{i0}^2 + \lambda \rho_{i1}^2, 
\end{align}
where
\begin{align*}
    \rho_{ig}^2 =&\ \int \sigma_{ig}(x) W_i^2(x) \frac{e(x)}{f_g(x)} E(\mathrm{d}x) < \infty, \ \ \sigma_{ig}(x) = m_{ig}(x) (1 - m_{ig}(x)). 
\end{align*}
Here, $f_0(x)$, $f_1(x)$, and $e(x)$ are the twice continuously differentiable density functions of the matching criterion for the reference group, focal group, and their averaged values, while $E(x)$ is their cumulative distribution function; and
\begin{align*}
    \lambda = \lim_{n_0, n_1 \rightarrow \infty} \frac{n_0}{n_0 + n_1} \in (0, 1).
\end{align*}
For more details, see \citet{srihera2010nonparametric}.

\paragraph{Variance estimation. } To estimate asymptotic variance~\eqref{eq:variance}, we propose 
\begin{align}
    \begin{split}\label{eq:sigma_new}
        \hat{\sigma}_i^2 = \ & \frac{1}{n_0 + n_1} \sum\limits_{p_0 \in \mathcal{P}_0} \hat{\sigma}_{i0}(\theta_{p_0})\left[\sum\limits_{k \in \mathcal{P}_0}\sum\limits_{l \in \mathcal{P}_1}W_i\left(\bar{X}_{kl}\right)W_{p_0i}\left(\bar{X}_{kl}\right)\right]^2 \\
        & + \frac{1}{n_0 + n_1} \sum\limits_{p_1 \in \mathcal{P}_1} \hat{\sigma}_{i1}(\theta_{p_1})\left[\sum\limits_{k \in \mathcal{P}_0}\sum\limits_{l \in \mathcal{P}_1}W_i\left(\bar{X}_{kl}\right)W_{p_1i}\left(\bar{X}_{kl}\right)\right]^2,
    \end{split}
\end{align}
which accounts for our binary-item setting, as it replaces squared residuals with estimated conditional variances 
\begin{align*}
    \hat{\sigma}_{ig}(\theta_{p_g}) =&\ \hat{m}_{ig}(\theta_{p_g}) \left(1 - \hat{m}_{ig}(\theta_{p_g})\right)
\end{align*}
in the original estimator proposed by \citet{srihera2010nonparametric}:
\begin{align*} 
    \hat{\sigma}_i^2 = \ & \frac{1}{n_0 + n_1} \sum\limits_{p_0 \in \mathcal{P}_0}\left(Y_{p_0i} - \hat{m}_{i0}(\theta_{p_0})\right)^2\left[\sum\limits_{k \in \mathcal{P}_0}\sum\limits_{l \in \mathcal{P}_1}W_i\left(\bar{X}_{kl}\right)W_{p_0i}\left(\bar{X}_{kl}\right)\right]^2 \\
    & + \frac{1}{n_0 + n_1} \sum\limits_{p_1 \in \mathcal{P}_1}\left(Y_{p_1i} - \hat{m}_{i1}(\theta_{p_1})\right)^2\left[\sum\limits_{k \in \mathcal{P}_0}\sum\limits_{l \in \mathcal{P}_1}W_i\left(\bar{X}_{kl}\right)W_{p_1i}\left(\bar{X}_{kl}\right)\right]^2. 
\end{align*}
This approach is more convenient, as it accounts for the binary nature of item responses. 

\paragraph{Asymptotic distribution. }Under the conditions specified above and assuming the null hypothesis holds, it can be shown that the test statistic \eqref{eq:nonparametric_test} normalised by $\hat{\sigma}_i$ specified in~\eqref{eq:sigma_new} asymptotically follows a standard normal distribution:
\begin{align*}
    \frac{\sqrt{N}\widehat{T}_i}{\hat{\sigma}_i} \gotodN \mathcal{N}(0, 1), \quad N = \frac{n_0 n_1}{n_0 + n_1},
\end{align*}
for details see \citet[][Theorems 1 and 2]{srihera2010nonparametric}.

\paragraph{Support size and computation. }The original approach evaluates the test statistic~\eqref{eq:nonparametric_test} over all $n_0 \cdot n_1$ averaged pairs, which may significantly slow down data manipulation in statistical software, making the proposed method time-consuming and memory-intensive, especially for larger sample sizes. To address this issue and improve efficiency, we propose and employ an alternative technique to calculate a common support: 
\begin{enumerate}
    \item[(1)] The common support is initially calculated as in the original approach.
    \item[(2)] The empirical weights of unique values of the averaged points are then computed.
    \item[(3)] A fixed-sized random sample is generated from the unique values of the common support using these weights. 
\end{enumerate}
This reduces computational burden while preserving the representativeness of both matching criterion distributions. When using the reduced support, it is important to note that the original size of the product (i.e., $n_0 \cdot n_1$) must be replaced by the size of the newly defined support set. 


\subsection{Weight function}

The choice of the weight function $W_i(\cdot)$ in the test statistic \eqref{eq:nonparametric_test} is crucial, as it can significantly influence the statistical power. In this study, we consider three strategies.  


\subsubsection{Fixed weights}

First, we consider uniform weighting: 
\begin{align}\label{eq:fixed}
    W_i(x) = 1 \quad \forall x.
\end{align}
This non-informative option is useful when no prior information about the nature of \gls{dif} is available. 


\subsubsection{Optimal weights}\label{sec:methodology:weights:optimal}

Second, we adapt an optimal weight function, derived in \citet[][Theorem 2]{srihera2010nonparametric}, to the case of binary data and for the comparison of the \glspl{irc}. This weight function is intended to maximise the local asymptotic power of the test. 

Under the local alternative hypothesis $m_{i0} = m_{i1} + \frac{cs_i}{N}, c \neq 0$, where $s_i$ is a difference function, the normalised test statistic~\eqref{eq:nonparametric_test} converges to the normal distribution:
\begin{align*}
    \frac{\sqrt{N} \widehat{T}_i}{\hat{\sigma}_i} \gotodN \mathcal{N}\left(\frac{\mu_i}{\sigma_i}, 1\right),
\end{align*}
where $\mu_i = -\int W_i(x)\left(m_{i0}(x) - m_{i1}(x)\right)E(\mathrm{d}x)$ and $\sigma_i^2$ is given in \eqref{eq:variance}. The asymptotic power is then given by
\begin{align}\label{eq:power}
    \mathsf{P}\left(\abs{\frac{\sqrt{N}\widehat{T}_i}{\hat{\sigma}_i}} \geq q_{1 - \frac{\alpha}{2}}\right) \simeq 1 - \phi\left(\frac{\mu_i}{\sigma_i} + q_{1 - \frac{\alpha}{2}}\right) + \phi\left(\frac{\mu_i}{\sigma_i} - q_{1 - \frac{\alpha}{2}}\right),
\end{align}
which is an increasing function of $\abs{\frac{\mu_i}{\sigma_i}}$. Thus, the weight function that maximises the asymptotic power \eqref{eq:power} is the one that maximises the term $\abs{\frac{\mu_i}{\sigma_i}}$. This is equivalent to maximising the term: 
\begin{align*}
    \frac{\mu_i^2}{\sigma_i^2} = \frac{\left[\int W_i(x)s_i(x)E(\mathrm{d}x)\right]^2}{\int \left[(1 - \lambda) \sigma_{i0}(x) \frac{e(x)}{f_0(x)} + \lambda \sigma_{i1}(x) \frac{e(x)}{f_1(x)} \right] W_i^2(x) E(\mathrm{d}x)},
\end{align*}
which yields
\begin{align*}
    W_i(x) = \frac{s_i(x)}{(1 - \lambda)\sigma_{i0}(x)\frac{e(x)}{f_{0}(x)} + \lambda \sigma_{i1}(x)\frac{e(x)}{f_1(x)}}.
\end{align*}

\begin{figure}[htb]
    \centering
    \includegraphics[width = \textwidth]{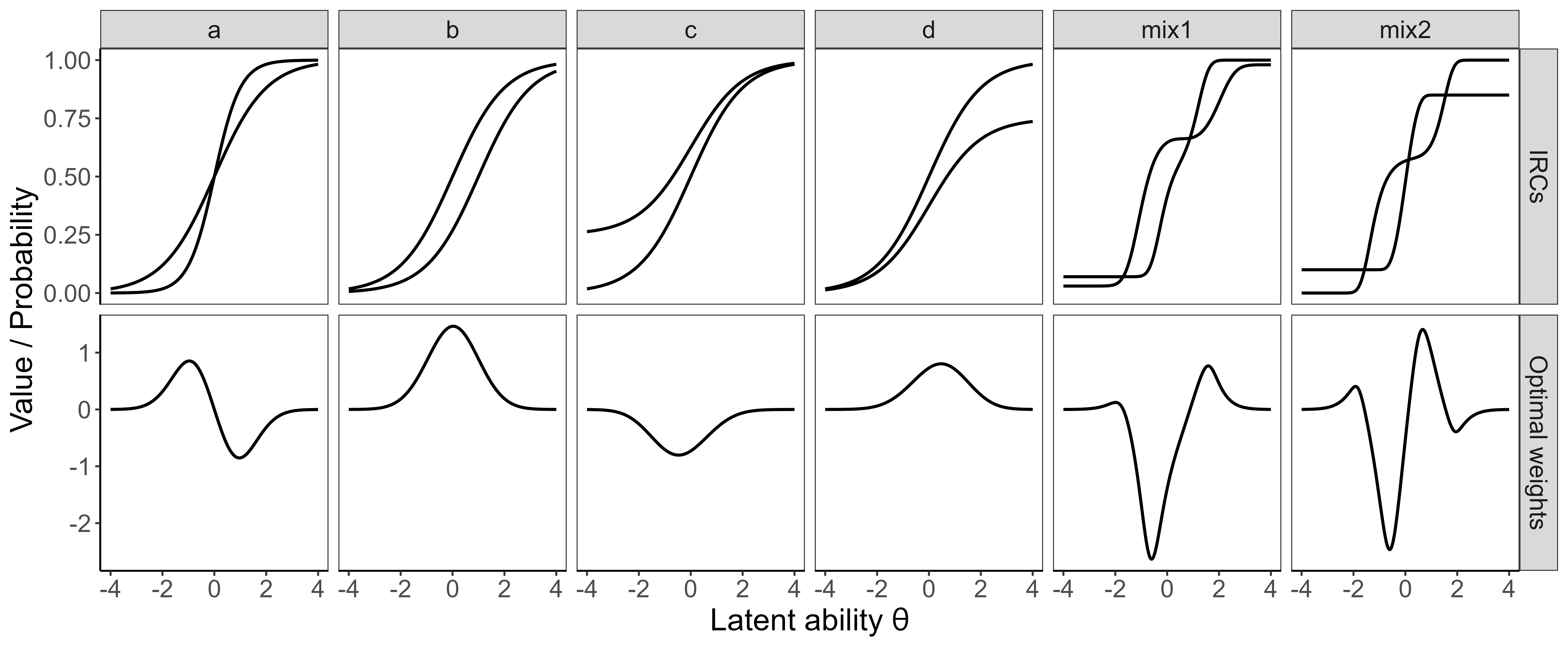}
    \caption[Examples of optimal weight functions]{Examples of IRCs and corresponding optimal weight functions \eqref{eq:optimal} for DIF caused by various parameters $a$, $b$, $c$, and $d$ in 4~\gls{pl} IRT model, and for logistic curves with several inflection points using normally distributed latent trait for both groups.}\label{fig:illustration_optimal_examples}
\end{figure}

Differences between the \glspl{irc} cannot generally be captured by a generic function, such as a polynomial. Therefore, in the context of this paper and its simulation study, we assume $s_i(x) = m_{i0}(x) - m_{i1}(x)$, representing the true difference between the two \glspl{irc}. Under this definition, the optimal weight function is given by 
\begin{align}\label{eq:optimal}
    W_i(x) = \frac{m_{i0}(x) - m_{i1}(x)}{(1 - \lambda) \sigma_{i0}(x) \frac{e(x)}{f_0(x)} + \lambda \sigma_{i1}(x) \frac{e(x)}{f_1(x)}}.
\end{align}
Figure~\ref{fig:illustration_optimal_examples} presents examples of \glspl{irc} showing \gls{dif} caused by different parameters, along with the corresponding optimal weights. Note that these weight functions can take negative values, enabling the detection of crossing non-uniform \gls{dif}, where \glspl{irc} intersect. Although the exact weights in~\eqref{eq:optimal} cannot be directly used in practice, since the true curve differences are unknown, they can serve as a valuable performance benchmark.

\subsubsection{Estimates of optimal weights}

Third, to make \eqref{eq:optimal} effective in practice, we extend the approach outlined in Section~\ref{sec:methodology:weights:optimal} by replacing the unknown quantities with their estimates. This yields a natural estimate of the optimal weights, 
\begin{align}\label{eq:optimal_estimate}
    \widehat{W}_i(x) = \frac{\hat{m}_{i0}(x) - \hat{m}_{i1}(x)}{(1 - \hat{\lambda})\hat{\sigma}_{i0}(x)\frac{\hat{e}(x)}{\hat{f}_0(x)} + \hat{\lambda}\hat{\sigma}_{i1}(x)\frac{\hat{e}(x)}{\hat{f}_1(x)}}.
\end{align}
Substituting $\widehat{W}_i(x)$ into \eqref{eq:nonparametric_test}, the resulting test statistic for item $i$ is given by:
\begin{align}\label{eq:nonparametric_estimate}
    \widehat{T}_i = \frac{1}{n_0 n_1}\sum_{p_0 \in \mathcal{P}_0}\sum_{p_1 \in \mathcal{P}_1} \frac{\left[\hat{m}_{i0}\left(\bar{X}_{p_0p_1}\right) - \hat{m}_{i1}\left(\bar{X}_{p_0p_1}\right)\right]^2}{(1 - \hat{\lambda})\hat{\sigma}_{i0}(\bar{X}_{p_0p_1})\frac{\hat{e}(\bar{X}_{p_0p_1})}{\hat{f}_0(\bar{X}_{p_0p_1})} + \hat{\lambda}\hat{\sigma}_{i1}(\bar{X}_{p_0p_1})\frac{\hat{e}(\bar{X}_{p_0p_1})}{\hat{f}_1(\bar{X}_{p_0p_1})}}. 
\end{align}

The test statistic \eqref{eq:nonparametric_estimate} includes the squared difference $(\hat{m}_{i0}(x) - \hat{m}_{i1}(x))^2$ in the numerator. In contrast to the test statistic in~\eqref{eq:nonparametric_test}, which reflects the weighted average of raw differences, this version represents the average squared discrepancy between \glspl{irc}. It is specifically designed to maximise sensitivity to complex \gls{dif} patterns, including cases where the curves intersect. 

\paragraph{Assessing significance. } Because substituting estimated weights invalidates the original asymptotic normality, we assess significance using a wild bootstrap \citep{wu1986jackknife, mammen1993bootstrap}. This method is particularly suitable when the data exhibits heteroskedasticity \cite[e.g.,][]{hardle1993comparing}, which aligns with the binary nature of the responses discussed in this work. This resampling scheme proceeds as follows:
\begin{enumerate}
    \item[(1)] \textbf{Initial Step:} Estimates of the \glspl{irc} are computed using \eqref{eq:estimate}. Then, the optimal weights are estimated with \eqref{eq:optimal_estimate}, and the \gls{dif} detection procedure is applied using the test statistic \eqref{eq:nonparametric_estimate}. 
    \item[(2)]\textbf{Bootstrap Sampling:} Under the null hypothesis (i.e., no \gls{dif}), a common \gls{irc} for both groups is estimated, and the corresponding fitted values $\left\{\hat{y}_{pi}\right\}_{p = 1}^n$ are computed. 
    \begin{enumerate}
        \item[(2a)] For each bootstrap run $b \in \left\{1, \dots, B\right\}$, where $B$ is the number of bootstrap samples, a bootstrap sample $y^{\ast}_{pib}$ is generated directly from Bernoulli distribution using fitted values  $\hat{y}_{pi}$, meaning:
        \begin{align*}
            y^{\ast}_{pib} \sim \text{Bernoulli}(\hat{y}_{pi}), 
        \end{align*}
        to account for the binary nature of the data. 
        \item[(2b)] For each bootstrap sample, the \gls{dif} detection procedure is applied as in the original sample in the initial step, resulting in a set of the test statistics $\left\{\widehat{T}_{ib}\right\}_{b = 1}^B$. 
    \end{enumerate}
    \item[(3)] \textbf{Final Step:} The set of the test statistics $\left\{\widehat{T}_{ib}\right\}_{b = 1}^B$ is compared to the test statistic of the original sample. A conclusion on \gls{dif} is made based on a two-sided $p$-value:
    \begin{align*}
        p\text{-value} = \frac{1}{B} \sum_{b = 1}^B \mathbf{1} [\widehat{T}_i < \widehat{T}_{ib}], 
    \end{align*}
    and the predefined level of significance.
\end{enumerate}


\section{Simulation study}\label{sec:simulation}

We conducted a Monte Carlo simulation study to evaluate the statistical properties of the proposed nonparametric \gls{dif} detection method \eqref{eq:nonparametric_test} under various conditions and to compare its performance with that of the well-established logistic regression approach \citep{swaminathan1990detecting}. Specifically, we examined type I error control, statistical power, and accuracy of estimated optimal weights. 


\subsection{Simulation design}

In this part, we describe the design of the simulation study, including the data generation process, the \gls{dif} detection procedures and their implementation, as well as the evaluation of the results. 


\subsubsection{Data and DIF generation}

Binary item responses were generated for the reference and focal groups from a logistic regression model extended with higher-order terms to allow for multiple inflexion points. For respondent $p$ and item $i$, the probability of a correct response was defined as
\begin{align}
    \begin{split}\label{eq:inflex}
         \Prob(Y_{pi} = 1|\theta_p) = c_{i} + (d_{i} - c_{i}) \frac{e^{a_{i} (\theta_p - b_{i} - e_{i} \theta_p^2 - f_{i} \theta_p^3 - g_{i} \theta_p^5)}}{1 + e^{a_{i} (\theta_p - b_{i} - e_{i} \theta_p^2 - f_{i} \theta_p^3 - g_{i} \theta_p^5)}},
    \end{split}
\end{align}
where $a_i$ denotes the discrimination, $b_i$ the difficulty $c_i$ the pseudo-guessing parameter, and $d_i$ inattention/slip parameter. The additional item parameters $(e_i, f_i, g_i)$ generate more complex \gls{irc} shapes with multiple inflexion points. Respondent abilities $\theta_p$ were drawn from a standard normal distribution in both groups. 

For non-\gls{dif} items, responses were generated using the true 4\gls{pl} \gls{irt} model \citep{barton1981upper} (i.e., model \ref{eq:inflex} with $(e_i, f_i, g_i) = (0, 0, 0)$). Other item parameters were identical across groups and drawn from normal distributions: Discrimination $a_{i} \sim \mathcal{N}(1.1, 0.3)$, difficulty $b_{i} \sim \mathcal{N}(0, 1.1)$, guessing $c_{i} \sim \mathcal{N}(0.2, 0.05)$, and inattention $d_{i} \sim \mathcal{N}(0.8, 0.05)$. 

To generate differentially functioning items, we considered six different sources of \gls{dif} in total: changes in discrimination $a_{i}$, difficulty $b_{i}$, guessing $c_{i}$, or inattention $d_{i}$, and two mixture conditions. In the first setting, called mix1, parameters were selected such that the \glspl{irc} intersect exactly once, while in the second setting, called mix2, they intersected twice. 

The magnitude of \gls{dif} was calibrated so that the weighted unsigned area measure between the two \glspl{irc} \citep{siebert2013differential} equalled 0.196, corresponding to a large effect size. The \glspl{irc} and the corresponding optimal weight functions for \gls{dif} items are illustrated in Figure~\ref{fig:illustration_optimal_examples}, and their parameters are summarised in Table~\ref{tab:parameters_DIF}. Each simulated test consisted of 20 items, including one DIF item (5\% prevalence).

The standardised total test score was used as the matching criterion $\theta_p$. Although discrete rather than continuous, it reflects common practice in applied \gls{dif} analyses.

The total sample sizes of $n =$ 50, 100, 200, 300, and 400 were selected, with both groups being equally sized. Each condition was replicated 1,000 times. 

\begin{table}[ht]
\caption{Item parameters used to generate DIF items}\label{tab:parameters_DIF}
\centering
\begin{tabular}{p{1.2cm}*{14}{@{\,}S@{\,}}}
  \toprule
  \vspace{-0.5em}DIF source\vspace{-4em} & \multicolumn{7}{c}{Reference group} & \multicolumn{7}{c}{Focal group}\\ \cmidrule(lr){2-8} \cmidrule(lr){9-15}
  & {$a$} & {$b$} & {$c$} & {$d$} & {$e$} & {$f$} & {$g$} & {$a$} & {$b$} & {$c$} & {$d$} & {$e$} & {$f$} & {$g$} \\ 
  \midrule
  $a$ & 0.42 & 0.00 & 0.00 & 1.00 & 0.00 & 0.00 & 0.00  & 2.00 & 0.00  & 0.00 & 1.00 & 0.00 & 0.00 & 0.00 \\ 
  $b$ & 1.00 & 0.00 & 0.00 & 1.00 & 0.00 & 0.00 & 0.00  & 1.00 & 1.00  & 0.00 & 1.00 & 0.00 & 0.00 & 0.00 \\ 
  $c$ & 1.00 & 0.00 & 0.00 & 1.00 & 0.00 & 0.00 & 0.00 & 1.00 & 0.00  & 0.39 & 1.00 & 0.00 & 0.00 & 0.00 \\ 
  $d$ & 1.00 & 0.00 & 0.00 & 0.61 & 0.00 & 0.00 & 0.00 & 1.00 & 0.00  & 0.00 & 1.00 & 0.00 & 0.00 & 0.00 \\ 
  mix1 & 1.90 & 0.28 & 0.07 & 1.00 & 1.00 & -0.70 & 0.00 & 0.35 & -1.75 & 0.03 & 0.98 & 1.60 & -0.90 & 0.00 \\ 
  mix2 & 4.20 & 0.00 & 0.10 & 0.85 & 0.00 & -0.50 & -0.50 & 0.18 & -1.50 & 0.00 & 1.00 & 1.00 & -1.20 & -0.50 \\
   \bottomrule
\end{tabular}%
\end{table}


\subsubsection{DIF detection}

Five approaches for \gls{dif} detection were evaluated in the simulation study: Four variations of the proposed nonparametric approach and the logistic regression method for \gls{dif} detection with the likelihood ratio test \citep{swaminathan1990detecting}. The nonparametric methods differed by the choice of the weight function: the fixed weights~\eqref{eq:fixed}, the theoretical optimal weights~\eqref{eq:optimal}, the estimated optimal weigh~\eqref{eq:optimal_estimate} without bootstrap calibration (i.e., assuming asymptotically normal distribution of the test statistic \eqref{eq:nonparametric_test}), and the estimate of optimal weights using wild bootstrap. The optimal weight function was applied only to \gls{dif} items, with values set to zero for non-\gls{dif} items, which necessarily yielded rejection rates of zero in the latter case. For the bootstrap-based method, the number of samples was set to $B = 500$. 

For kernel estimation of \glspl{irc}, the Epanechnikov kernel was used with three different bandwidth parameters $h = n_0^{-\zeta}$, where $\zeta$ took values $0.260, \frac{7}{24} (\approx 0.292)$, and $0.320$,
satisfying the regularity conditions. All tests are performed at a significance level of 0.05.


\subsubsection{Evaluation of the results}

The five different \gls{dif} detection techniques (four variations of the nonparametric approach and the logistic regression method) were compared on two key performance metrics: power and rejection rate. Power is defined as the proportion of true positives (i.e., correctly detected \gls{dif} items), while rejection rate refers to the proportion of false positives (i.e., non-\gls{dif} items incorrectly identified as \gls{dif}). 

Additionally, the accuracy of the estimate of the optimal weights was assessed by computing the \gls{rmse}, which quantifies the root of the mean squared difference between the optimal weights \eqref{eq:optimal} and their estimates. 


\subsubsection{Implementation}

All analyses were conducted in the statistical software \pkg{R} \citep{R2019}, version 4.3.2, and its associated packages. Empirical density functions were computed with the \Rcode{ecdf()} function from the \pkg{stats} package \citep{R2019}. Weights of kernel functions and kernel estimates were calculated by the \Rcode{locCteWeightsC()} and \Rcode{locWeightsEval()} functions from the \CRANpkg{locpol} package \citep{cabrera2018locpol}. Estimates of densities of standardised total scores and the common support of the test statistic were evaluated with the \Rcode{bkde()} function from the \CRANpkg{KernSmooth} package \citep{wand2019kernsmooth}. The logistic regression method with the likelihood ratio test was performed using the \Rcode{glm()} function from the \CRANpkg{stats} package. Finally, graphical representations of the results were created using the \CRANpkg{ggplot2} package \citep{wickham2016ggplot2}.


\subsection{Simulation results}


\subsubsection{Rejection rates and power}

The estimates of the optimal weights \eqref{eq:estimate} without the wild bootstrap were the most powerful approach across all scenarios, with a mean power rate of 0.724. However, this gain in sensitivity came at the cost of a substantially inflated Type I error, with an average rejection rate of 0.272, exceeding the nominal significance level of 0.05. Consequently, this variant was considered unreliable and was excluded from subsequent analyses; its results are therefore not reported here. 

\begin{figure}[hbt]
    \centering
    \includegraphics[width = 0.985\textwidth]{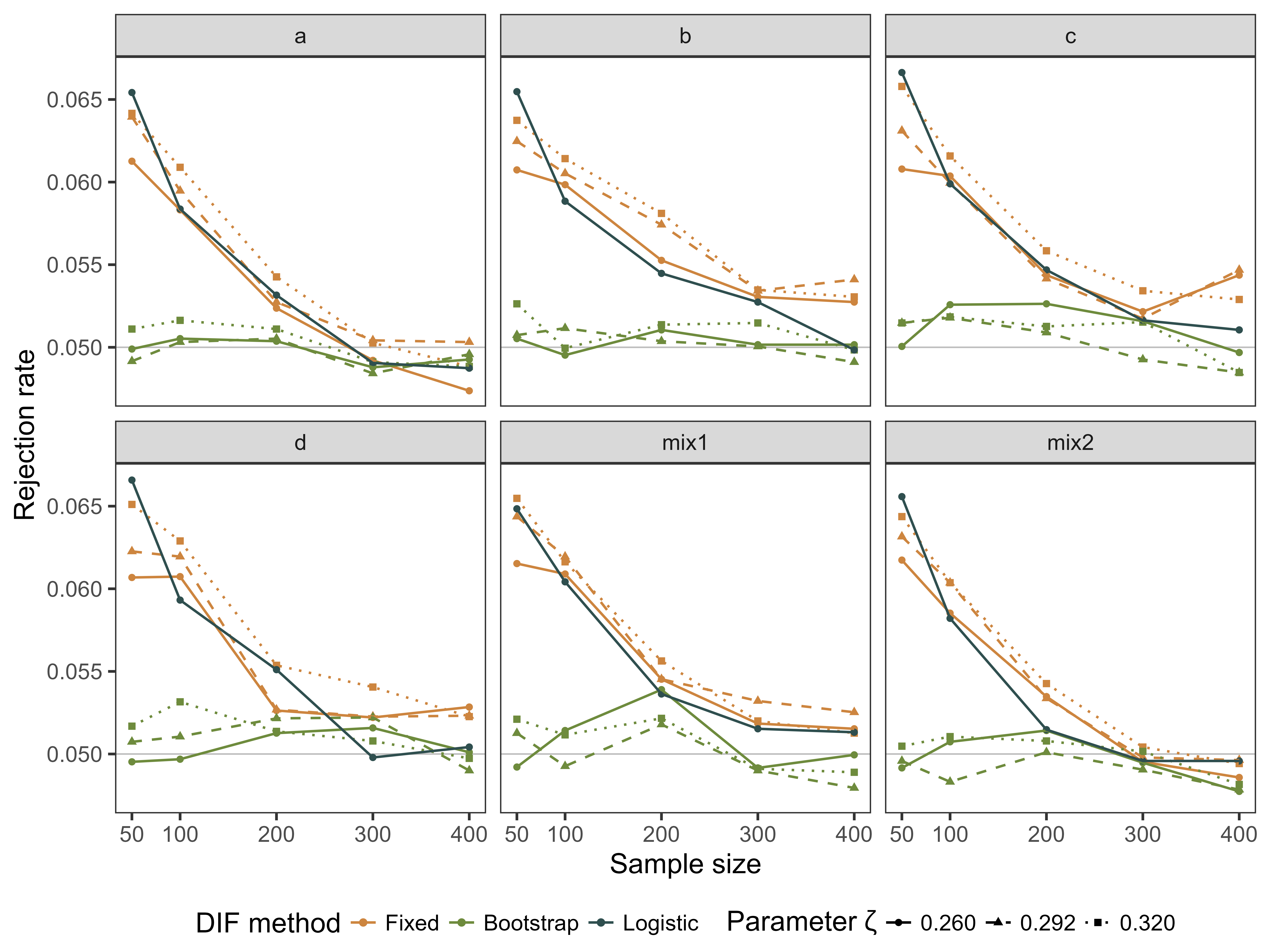}
    \caption[Rejection rates]{Rejection rates by the nonparametric approach with various weight functions and by the logistic regression method with respect to the sample size and the parameter $\zeta$ for different sources of DIF; the horizontal line shows a significance level of 0.05.}\label{fig:simulation_RR}
\end{figure}

All remaining approaches maintained appropriate control of the significance level across all bandwidth parameters $\zeta$, \gls{dif} sources, and sample sizes, with rejection rates ranging from 0.050 to 0.069. No significant differences were observed in rejection rates among the \gls{dif} detection methods, regardless of \gls{dif} sources. A slight increase in rejection rates was noted for both the logistic regression method and the nonparametric approach with fixed weights at the smallest sample size ($n = 50$; Figure~\ref{fig:simulation_RR}, Table \ref{app:tab:simulation_RR}). 

\begin{figure}[hbt]
    \centering
    \includegraphics[width = 0.985\textwidth]{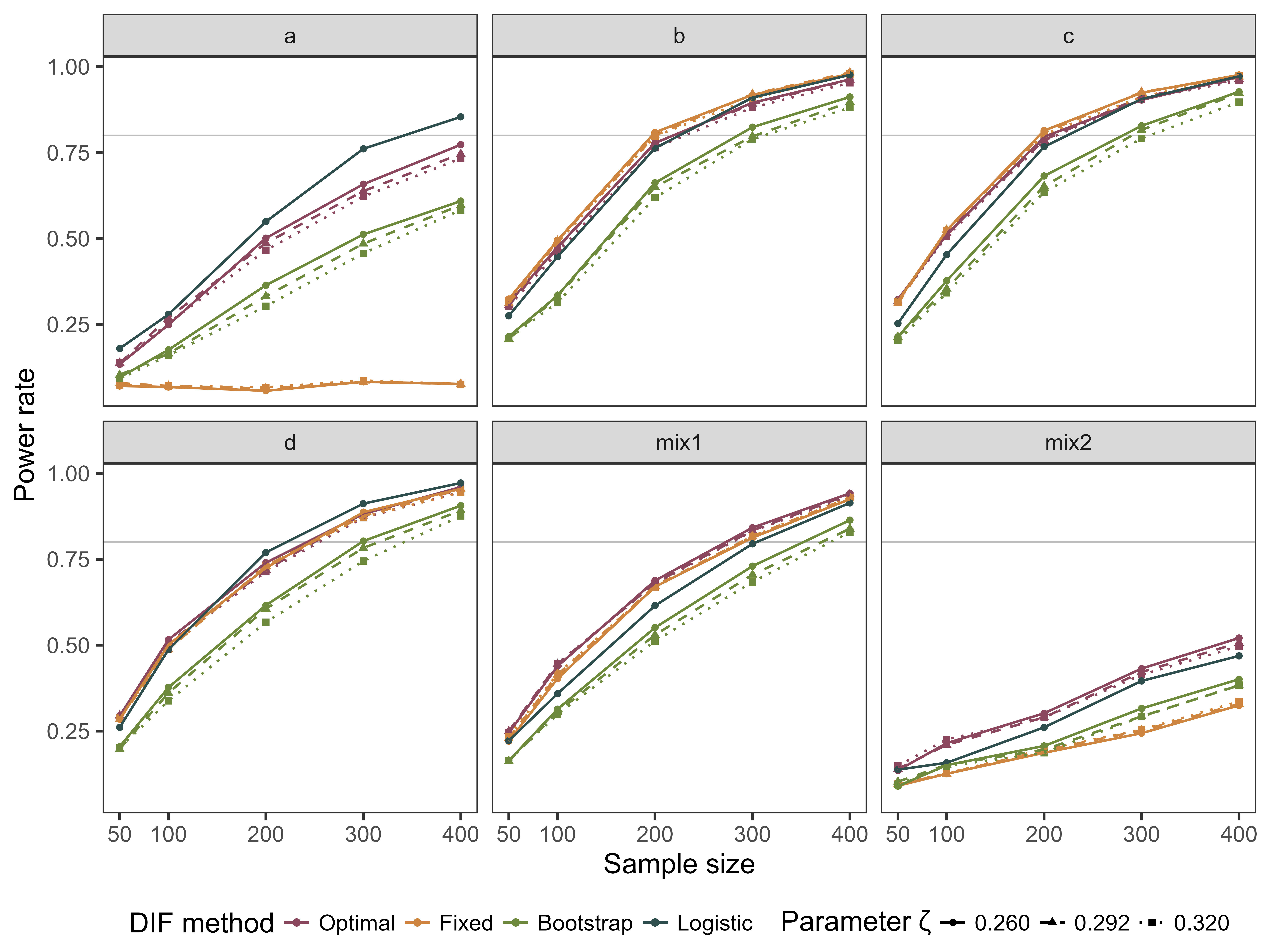}
    \caption[Power rates]{Power rates by nonparametric approach with various weight functions and by the logistic regression method with respect to the sample size and the parameter $\zeta$ for different sources of DIF; the horizontal line shows sufficient power of 0.80.}\label{fig:simulation_PR}
\end{figure}

All \gls{dif} detection methods exhibited lower power at smaller sample sizes, as expected. With increasing sample size, power improved for all methods, and differences between approaches became less pronounced (Figure~\ref{fig:simulation_PR}, Table~\ref{app:tab:simulation_PR}). The nonparametric approaches, using either optimal or fixed weights, outperformed the logistic regression method in scenarios where shifts in parameters $b$ and $c$ were sources of \gls{dif}, as well as in the mix1 scenario, across nearly all sample sizes. This was also the case for small samples ($n \leq 100$) when \gls{dif} was caused by parameter $d$. Furthermore, in the mix2 scenario, the nonparametric approach using optimal weights achieved higher power than logistic regression. In contrast, logistic regression gained the highest power when parameter $a$ was the source of \gls{dif}. In this case, the nonparametric approach with the fixed weights failed to detect \gls{dif} effectively and achieved only limited power, also in the mix2 scenario. Under such circumstances, the nonparametric method with estimated optimal weights and wild bootstrap provided a marked improvement, consistently yielding higher power across all sample sizes. 

Generally, the differences between the nonparametric approaches using different bandwidth parameters $h$ were small. When the value of $\zeta$ was lower, meaning the bandwidth parameter $h = n_0^{-\zeta}$ was larger, the optimal weights and their estimates using the wild bootstrap yielded slightly higher mean power. 


\subsubsection{Estimates of optimal weights}

In nine scenarios where the parameter $c$ was the source of \gls{dif} and $\zeta = 0.292$, the estimates of the optimal weights exceeded $10^6$, indicating numerical instability or divergence. As a result, these scenarios were excluded from further analysis to ensure the reliability of findings and are not reported here. 

\begin{figure}[hbt]
    \centering
    \includegraphics[width = 0.985\textwidth]{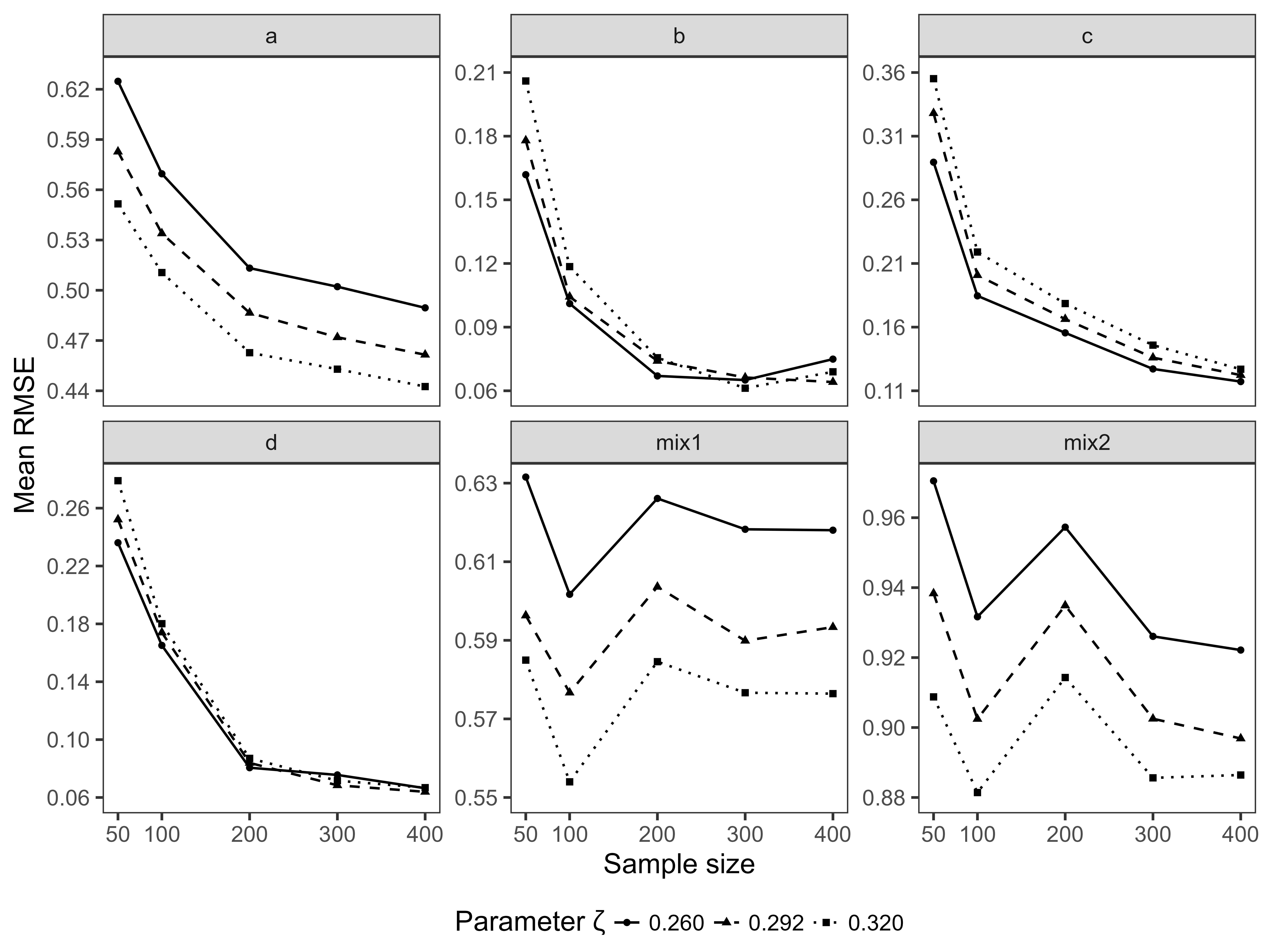}
    \caption{RMSE of the estimates of optimal weights with respect to the parameter $\zeta$, source of DIF, and sample size.}
    \label{fig:simulation_MSE}
\end{figure}

Estimation of the optimal weight was the most precise when the parameters $b$, $c$, and $d$ were sources of \gls{dif}. In contrast, it was the most biased when the mix2 scenario was considered (Figure~\ref{fig:simulation_MSE}, Table \ref{app:tab:simulation_MSE}; see also rows 2, 3, 4, and 6 in Figure~\ref{app:fig:simulation_weights}). The smallest overall \gls{rmse} of 0.242 was achieved for $\zeta = 0.320$ (the smallest bandwidth parameter $h$), while the largest overall \gls{rmse} of 0.270 occurred for $\zeta = 0.260$ (the largest $h$).

All three choices of the $\zeta$ parameter resulted in more accurate estimates for larger sample sizes compared to smaller ones. While there were no significant differences between bandwidth parameters when the parameters $b$, $c$, or $d$ were sources of \gls{dif}, this was not the case when the \glspl{irc} intersected. In such scenarios, $\zeta = 0.320$ (the smallest bandwidth parameter) produced the most accurate estimates, whereas $\zeta = 0.260$ (the largest bandwidth parameter) consistently resulted in the highest \gls{rmse} across all sample sizes. Furthermore, the precision of the estimates did not always increase (i.e., decrease of \gls{rmse}) with increasing sample size when the discrimination parameter $a$ was a source of \gls{dif} or when the mix1 setting was considered for the underlying \glspl{irc} (Figure~\ref{fig:simulation_MSE}, Table \ref{app:tab:simulation_MSE}; see also rows 1 and 5 in Figure~\ref{app:fig:simulation_weights}). 


\section{Real data example}\label{sec:real_data}

To illustrate the practical application of our methodology, we analyse data from a questionnaire on verbal aggression. 

\subsection{Data description}

The Verbal Aggression dataset 
(\citeauthor{vansteelandt2001formal}, \citeyear{vansteelandt2001formal}; available in \citeauthor{magis2010general}, \citeyear{magis2010general}) contains responses of 316 participants (243 females and 73 males) to a 24-item questionnaire measuring tendencies toward verbal aggression. Each item describes a frustrating situation paired with a potential verbal aggression reaction. Specifically, four frustration situations were considered: (S1) A bus fails to stop for me; (S2) I miss a train because a clerk gave me faulty information; (S3) The grocery store closes just as I am about to enter; (S4) The operator disconnects me when I had used up my last 10 cents for a call. For each situation, respondents indicated whether they wanted to or actually did react with one of three possible aggressive behaviours: cursing, scolding, or shouting. For instance, item \textit{S1WantShout} corresponds to the statement "A bus fails to stop for me. I want to shout". Items are binary-coded. A response of 1 denotes agreement with the statement, and 0 indicates disagreement.

\subsection{Statistical analysis}

Three approaches for \gls{dif} detection were evaluated in the real data example analysis: Two variations of the proposed nonparametric approach and the logistic regression method for \gls{dif} detection with the likelihood ratio test \citep{swaminathan1990detecting}. The nonparametric methods differed in the choice of the weight function: the fixed weights~\eqref{eq:fixed} and the estimated optimal weights~\eqref{eq:estimate} using the bootstrap. For kernel estimation of \glspl{irc}, the Epanechnikov kernel was used with three different bandwidth parameters $h = n_0^{-\zeta}$, where $\zeta$ took values $0.260, \frac{7}{24} (\approx 0.292)$, and $0.320$. \gls{dif} was analysed with respect to the respondent gender. A standardised total score was used as the measure of observed matching ability. All tests are performed at a significance level of 0.05.

\subsection{Results}

In the Verbal Aggression dataset, eleven items were flagged as exhibiting \gls{dif} by at least one of the detection methods (Table~\ref{tab:verbal:stats}). Three items (\textit{S2WantShout}, \textit{S2DoCurse}, and \textit{S2DoScold}) were consistently identified across all approaches. By contrast, the item \textit{S1WantScold} was detected solely by the nonparametric method with estimated optimal weights using bootstrap. 

Across the nonparametric variants, the fixed-weight method and the estimated optimal weights with bandwidth $\zeta = 0.32$ identified the largest number of \gls{dif} items (seven each). The versions with 
$\zeta = 0.292$ and $\zeta = 0.26$ followed, flagging six and five items, respectively. Logistic regression identified the fewest \gls{dif} items, with only four flagged.

Alignment between methods varied. The highest agreement was observed within the fixed-weight variants (tetrachoric correlations between 0.90 and 0.99) and within the estimated optimal weights using bootstrap (0.89--0.97). Agreement between these two nonparametric families was lower, with correlations ranging from 0.33 to 0.66. Logistic regression showed moderate alignment with nonparametric approaches, correlating 0.86--0.87 with the fixed-weight methods and 0.71--0.84 with the estimated optimal weight using bootstrap.

\begin{table}[ht]
\centering
\caption{Test statistics and p-values for DIF items from the Verbal Aggression dataset.} 
\label{tab:verbal:stats}
 \resizebox{\textwidth}{!}{%
\begin{tabular}{lrrrrrrrrrrrrrr}
  \toprule
  \multicolumn{1}{c}{\multirow{4}{*}{Item}} & \multicolumn{12}{c}{Nonparametric} & \multicolumn{2}{c}{\multirow{3}{*}{Logistic}} \\ \cmidrule(lr){2-13}  
  & \multicolumn{6}{c}{Fixed} & \multicolumn{6}{c}{Bootstrap} & & \\ \cmidrule(lr){2-7} \cmidrule(lr){8-13} 
  &  \multicolumn{2}{c}{$\zeta = 0.260$} & \multicolumn{2}{c}{$\zeta = 0.292$} & \multicolumn{2}{c}{$\zeta = 0.320$} & \multicolumn{2}{c}{$\zeta = 0.260$} & \multicolumn{2}{c}{$\zeta = 0.292$} & \multicolumn{2}{c}{$\zeta = 0.320$}  & & \\ \cmidrule(lr){2-3} \cmidrule(lr){4-5} \cmidrule(lr){6-7} \cmidrule(lr){8-9} \cmidrule(lr){10-11} \cmidrule(lr){12-13} \cmidrule(lr){14-15} 
  &  $T$-value & $p$-value & $T$-value & $p$-value & $T$-value & $p$-value & $T$-value & $p$-value & $T$-value & $p$-value & $T$-value & $p$-value & $\chi^2$-value & $p$-value \\ 
 \midrule
S1WantScold & $1.369$ & $0.171$\hphantom{*} & $1.313$ & $0.189$\hphantom{*} & $1.280$ & $0.201$\hphantom{*} & $2.135$ & $0.144$\hphantom{*} & $2.491$ & $0.060$\hphantom{*} & $2.936$ & $0.030$* & $3.354$ & $0.187$\hphantom{*} \\ 
  S2WantCurse & $1.667$ & $0.096$\hphantom{*} & $1.694$ & $0.090$\hphantom{*} & $1.737$ & $0.082$\hphantom{*} & $1.649$ & $0.000$* & $1.602$ & $0.000$* & $1.780$ & $0.000$* & $4.730$ & $0.094$\hphantom{*} \\ 
  S2WantScold & $1.914$ & $0.056$\hphantom{*} & $2.114$ & $0.034$* & $1.969$ & $0.049$* & $2.471$ & $0.072$\hphantom{*} & $2.558$ & $0.090$\hphantom{*} & $2.740$ & $0.070$\hphantom{*} & $4.140$ & $0.126$\hphantom{*} \\ 
  S2WantShout & $3.204$ & $0.001$* & $3.358$ & $0.001$* & $3.369$ & $0.001$* & $3.564$ & $0.002$* & $3.355$ & $0.004$* & $3.526$ & $0.000$* & $11.411$ & $0.003$* \\ 
  S4WantShout & $2.098$ & $0.036$* & $2.327$ & $0.020$* & $2.216$ & $0.027$* & $2.503$ & $0.058$\hphantom{*} & $2.490$ & $0.082$\hphantom{*} & $2.580$ & $0.106$\hphantom{*} & $3.688$ & $0.158$\hphantom{*} \\ 
  S1DoScold & $-2.128$ & $0.033$* & $-1.949$ & $0.051$\hphantom{*} & $-1.848$ & $0.065$\hphantom{*} & $2.172$ & $0.136$\hphantom{*} & $2.078$ & $0.000$* & $2.110$ & $0.000$* & $4.730$ & $0.094$\hphantom{*} \\ 
  S2DoCurse & $-2.964$ & $0.003$* & $-2.946$ & $0.003$* & $-2.966$ & $0.003$* & $3.064$ & $0.026$* & $3.066$ & $0.034$* & $3.514$ & $0.024$* & $7.693$ & $0.021$* \\ 
  S2DoScold & $-3.050$ & $0.002$* & $-2.871$ & $0.004$* & $-2.742$ & $0.006$* & $3.045$ & $0.012$* & $3.012$ & $0.014$* & $3.073$ & $0.024$* & $10.262$ & $0.006$* \\ 
  S2DoShout & $-0.493$ & $0.622$\hphantom{*} & $-0.130$ & $0.897$\hphantom{*} & $-0.070$ & $0.944$\hphantom{*} & $2.788$ & $0.030$* & $2.844$ & $0.024$* & $2.905$ & $0.028$* & $1.702$ & $0.427$\hphantom{*} \\ 
  S3DoCurse & $-2.686$ & $0.007$* & $-2.573$ & $0.010$* & $-2.667$ & $0.008$* & $2.596$ & $0.068$\hphantom{*} & $2.618$ & $0.076$\hphantom{*} & $2.650$ & $0.074$\hphantom{*} & $7.238$ & $0.027$* \\ 
  S3DoScold & $-2.112$ & $0.035$* & $-2.134$ & $0.033$* & $-2.141$ & $0.032$* & $2.153$ & $0.100$\hphantom{*} & $2.118$ & $0.188$\hphantom{*} & $2.242$ & $0.158$\hphantom{*} & $5.868$ & $0.053$\hphantom{*} \\ 
   \bottomrule
\end{tabular}
}
\end{table}

Across all items, the nonparametric approach with the bandwidth parameter $\zeta = 0.32$ produced the most accurate estimates of the \glspl{irc}, with the exception of \textit{S2DoScold}, \textit{S3DoShout}, and \textit{S4DoShout}, where the logistic regression model achieved slightly lower squared bias. Conversely, logistic regression yielded the least precise estimates for 14 of the 24 items, while the nonparametric method with 
$\zeta = 0.26$ showed the lowest precision for the remaining 10 items.

\begin{figure}[h]
    \centering
    \begin{subfigure}[t]{.495\textwidth}
        \centering
        \includegraphics[width = 0.9\textwidth]{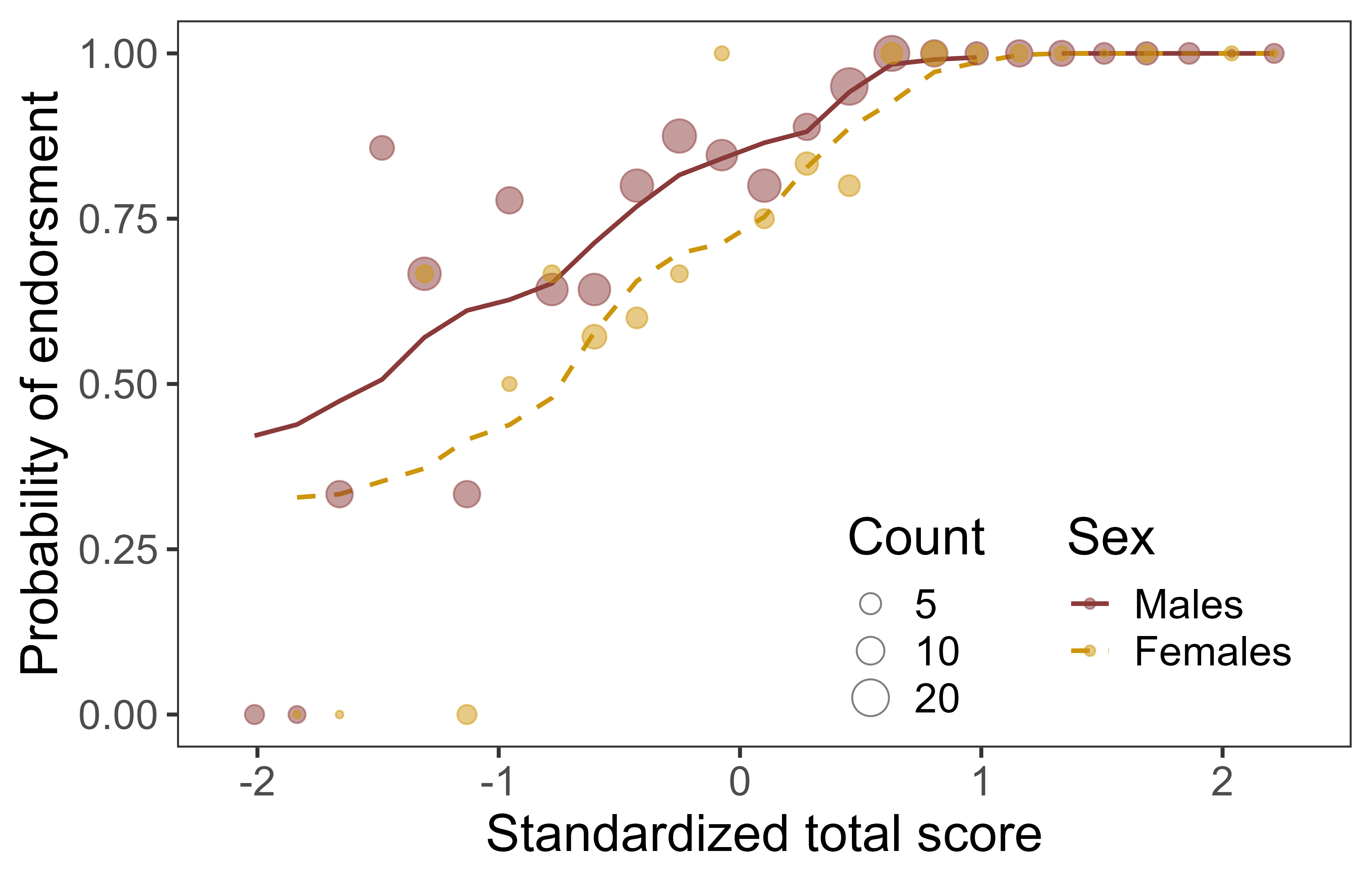}
        \caption{Nonparametric approach with $\zeta = 0.32$}\label{fig:verbal:nonparametric}
    \end{subfigure}
    \begin{subfigure}[t]{.495\textwidth}
        \centering
        \includegraphics[width = 0.9\textwidth]{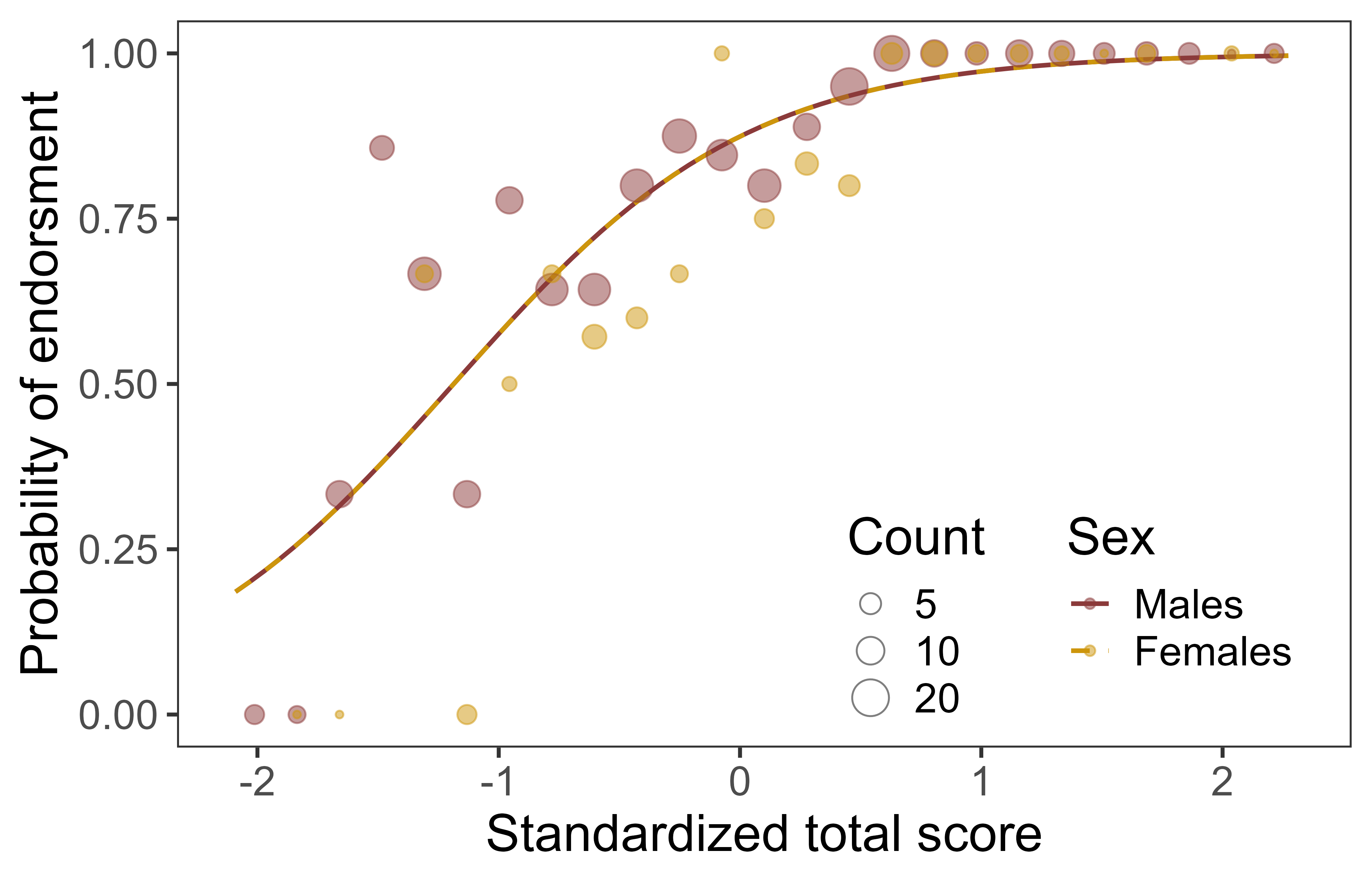}
        \caption{Logistic regression}\label{fig:verbal:logistic}
  \end{subfigure}
    
    \caption{IRCs of the \textit{S2WantCurse} item by the nonparametric approach with $\zeta = 0.32$ and logistic regression. }
    \label{fig:verbal:icc:S2WantCurse}
\end{figure}

To illustrate how the proposed method performs in practice, the item \textit{S2WantCurse} provides a representative example. It was estimated most precisely by the nonparametric approach with the bandwidth $\zeta = 0.32$ and was also identified as exhibiting \gls{dif} by this method (Figure~\ref{fig:verbal:nonparametric}). In contrast, logistic regression produced a visibly poorer fit for the same item and did not classify it as \gls{dif} (Figure~\ref{fig:verbal:logistic}). This example demonstrates the ability of the nonparametric framework to reveal subtle group differences that may be undetected under parametric modelling.

\section{Discussion}\label{sec:discussion}

In this work, we proposed a novel nonparametric approach for comparing \glspl{irc} to detect \gls{dif} in binary items. Building on the general framework for comparing regression curves introduced by \citet{srihera2010nonparametric}, we adapted their methodology to address a common challenge in psychometrics and social sciences: testing differences between \glspl{irc}. Our main methodological contributions include (1) a new estimator of the asymptotic variance of the test statistic tailored to account for the binary nature of the data, (2) a derivation of optimal weight functions that maximize the local power of the test, and a procedure for estimating these weights in a realistic setting where they are unknown, and (3) a wild bootstrap procedure to approximate the unknown asymptotic distribution of the test statistic when using estimate of optimal weight, enabling robust hypothesis testing. These innovations extend existing approaches by providing a flexible, practical, and theoretically grounded framework for \gls{dif} detection that directly estimates \glspl{irc}, allowing nuanced detection of group differences that may be missed by parametric methods.

To evaluate the performance of the proposed approach, we conducted a Monte Carlo simulation study comparing it using various weighting schemes to the logistic regression method. All methods demonstrated good control of Type I error. The nonparametric approach using the optimal weights achieved power rates close to those of the logistic regression method, and it outperformed it in several scenarios, especially in scenarios with multiple intersections of the underlying \glspl{irc}. When comparing different weight functions within the nonparametric approach, the fixed weights performed similarly to the optimal weights in cases where the \glspl{irc} did not intersect, and may be recommended when it can be assumed that one group is advantaged over the other group for all levels of the matching criterion. However, when \glspl{irc} intersected, the estimate of the optimal weights using the wild bootstrap technique substantially improved performance over fixed weights. 


To illustrate the proposed \gls{dif} detection method, we analysed a real-life dataset on verbal aggression. Across methods, eleven items were flagged for \gls{dif}, with three items consistently detected by all approaches. The nonparametric methods, especially those with fixed weights and estimated optimal weights with bandwidth $\zeta = 0.32$, identified the most \gls{dif} items, while logistic regression flagged the fewest. Agreement among nonparametric variants was high, but alignment between nonparametric methods and the logistic regression method was moderate. Importantly, the nonparametric method with $\zeta = 0.32$ generally produced the most accurate \gls{irc} estimates and successfully detected subtle \gls{dif} undetected by logistic regression. This demonstrates the potential of the proposed nonparametric framework to uncover nuanced group differences that parametric models may overlook.

Our approach complements existing nonparametric approaches, such as the Mantel-Haenszel test \citep{mantel1959statistical, holland1988differential}, the SIBTEST method \citep{shealy1993model}, or standardisation \citep{dorans1986demonstrating}, which do not directly model \glspl{irc}. In contrast, the newly proposed method explicitly estimates and compares \glspl{irc}, allowing for flexible, data-driven weighting that enhances sensitivity to complex \gls{dif} patterns, including intersections and nonuniform differences between groups. It also complements kernel smoothing \gls{dif} detection methods, such as a kernel-smoothed SIBTEST \citep{douglas1996kernel} or TestGraf, a graphical \gls{dif} method with a kernel smoothing for estimating the conditional probability of correct answers related to proficiency estimates \citep{bolt2006testing, ramsay2000testgraf}. Unlike these approaches, our method is grounded in a direct comparison of \glspl{irc} within a unified nonparametric framework, offering both interpretability and methodological flexibility.

In past decades, many authors have dealt with the topic of nonparametric comparison of regression curves, including \citet{dette2001nonparametric, hall1990bootstrap, neumeyer2003nonparametric} and \citet{scheike2000comparison}. Our work builds on the general approach of \citet{srihera2010nonparametric}, which accommodates a random design and therefore allows for direct extension addressing challenges specific to \gls{dif} detection in real-life settings of binary items.  

The current study has several limitations, as well as potential directions for future research. 
First, the simulation study was limited in terms of the number of items, the proportion of \gls{dif} items, and sample sizes, as only small to moderate sample sizes were considered to ensure computational feasibility. This limitation precluded the inclusion of extended logistic regression models such as 3\gls{pl} or 4\gls{pl} models \citep{barton1981upper, birnbaum1968statistical, hladka2025estimation} in the simulation study, as they require larger sample sizes for both groups. 
Second, this study focused exclusively on the Epanechnikov kernel. While alternative kernel functions could be considered, previous research suggests that the accuracy of estimation is generally robust to the choice \citep{douglas1996kernel}. Nonetheless, future studies could investigate whether alternative kernels offer practical advantages for detecting \gls{dif} in various settings.
Third, three levels of the bandwidth parameter $h$ were examined. The choice of bandwidth is directly related to the precision of estimating optimal weight functions: If the $h$ is too small, the resulting estimate may be under-smoothed, leading to high variance. Conversely, a large $h$ may result in over-smoothing, which can obscure important features of the data. The bandwidth values chosen for this study were intended to cover a plausible range, while no large differences in power or rejection rates were observed.
Fourth, the kernel-smoothing estimate of \glspl{irc} does not require monotonicity of item responses, which is a typical assumption in logistic regression or \gls{irt} models.  Parametric models typically assume that the probability of a correct response increases monotonically with the latent trait. This assumption simplifies estimation and interpretation but may not always hold in practice, particularly when items are affected by multidimensional traits, guessing effects, or complex DIF patterns. However, the monotonicity assumption ensures the interpretability and scalability of test scores \cite[see, e.g.,][]{mokken1971theory}, required, for example, in nonparametric \gls{irt} models \citep{douglas2001asymptotic, he2024extended}.

In summary, the proposed nonparametric approaches, including a novel estimate of the optimal weights with the wild bootstrap, demonstrated control of significance levels and, in most cases, matched or exceeded the performance of the logistic regression method in detecting \gls{dif}. Importantly, the flexibility of the nonparametric framework allows it to capture complex patterns in \glspl{irc}, particularly in scenarios involving multiple intersections or non-monotonic structures, where traditional parametric methods may falter. These results highlight the substantial potential of our approach as a powerful and universal tool for \gls{dif} detection, expanding methodological options for applied psychometrics and advancing the analysis of item functioning in real-world testing contexts. 


\bibliographystyle{abbrvnat}
\bibliography{references}


\clearpage

\begin{appendices}


\section{Tables}

\setcounter{table}{0}
\renewcommand{\thetable}{A\arabic{table}}

\begin{table}[ht]
\caption{Rejection rates by the nonparametric approach with various weight functions and by the logistic regression method with respect to the sample size $n$ and the parameter $\zeta$ for different sources of DIF.} 
\label{app:tab:simulation_RR}
\centering
\begin{tabular}{lr rrr rrr rrr r}
  \toprule
  \multirow{3}{*}{\shortstack{\\[0.5ex]DIF\\[0.75ex] source}} & \multirow{3}{*}{$n$} & \multicolumn{9}{c}{Nonparametric} & \multirow{3}{*}{Logistic} \\ \cmidrule(lr){3-11}
   &  & \multicolumn{3}{c}{Optimal} & \multicolumn{3}{c}{Fixed} & \multicolumn{3}{c}{Bootstrap} &  \\ \cmidrule(lr){3-5} \cmidrule(lr){6-8} \cmidrule(lr){9-11}
 & & 0.260 & 0.292 & 0.320 & 0.260 & 0.292 & 0.320 & 0.260 & 0.292 & 0.320 & \\ 
  \midrule
$a$ \\
  & 50 & 0.000 & 0.000 & 0.000 & 0.063 & 0.065 & 0.066 & 0.052 & 0.053 & 0.054 & 0.069 \\ 
  & 100 & 0.000 & 0.000 & 0.000 & 0.055 & 0.057 & 0.058 & 0.051 & 0.051 & 0.052 & 0.057 \\ 
  & 200 & 0.000 & 0.000 & 0.000 & 0.056 & 0.058 & 0.060 & 0.055 & 0.057 & 0.056 & 0.054 \\ 
  & 300 & 0.000 & 0.000 & 0.000 & 0.051 & 0.051 & 0.051 & 0.053 & 0.052 & 0.052 & 0.050 \\ 
  & 400 & 0.000 & 0.000 & 0.000 & 0.052 & 0.052 & 0.052 & 0.060 & 0.057 & 0.052 & 0.052 \\ 
$b$ \\
  & 50 & 0.000 & 0.000 & 0.000 & 0.065 & 0.066 & 0.069 & 0.056 & 0.059 & 0.057 & 0.068 \\ 
  & 100 & 0.000 & 0.000 & 0.000 & 0.059 & 0.060 & 0.061 & 0.057 & 0.056 & 0.056 & 0.062 \\ 
  & 200 & 0.000 & 0.000 & 0.000 & 0.054 & 0.056 & 0.057 & 0.057 & 0.056 & 0.057 & 0.054 \\ 
  & 300 & 0.000 & 0.000 & 0.000 & 0.056 & 0.057 & 0.058 & 0.054 & 0.056 & 0.051 & 0.055 \\ 
  & 400 & 0.000 & 0.000 & 0.000 & 0.056 & 0.056 & 0.057 & 0.058 & 0.057 & 0.053 & 0.056 \\
$c$ \\
  & 50 & 0.000 & 0.000 & 0.000 & 0.062 & 0.063 & 0.066 & 0.053 & 0.055 & 0.056 & 0.066 \\ 
  & 100 & 0.000 & 0.000 & 0.000 & 0.056 & 0.057 & 0.059 & 0.053 & 0.055 & 0.054 & 0.056 \\ 
  & 200 & 0.000 & 0.000 & 0.000 & 0.055 & 0.055 & 0.057 & 0.052 & 0.054 & 0.053 & 0.055 \\ 
  & 300 & 0.000 & 0.000 & 0.000 & 0.054 & 0.056 & 0.059 & 0.057 & 0.057 & 0.055 & 0.053 \\ 
  & 400 & 0.000 & 0.000 & 0.000 & 0.058 & 0.058 & 0.060 & 0.058 & 0.056 & 0.054 & 0.055 \\ 
$d$ \\
  & 50 & 0.000 & 0.000 & 0.000 & 0.067 & 0.067 & 0.069 & 0.055 & 0.054 & 0.056 & 0.069 \\ 
  & 100 & 0.000 & 0.000 & 0.000 & 0.056 & 0.056 & 0.057 & 0.054 & 0.053 & 0.055 & 0.055 \\ 
  & 200 & 0.000 & 0.000 & 0.000 & 0.054 & 0.055 & 0.055 & 0.055 & 0.055 & 0.053 & 0.053 \\ 
  & 300 & 0.000 & 0.000 & 0.000 & 0.057 & 0.058 & 0.057 & 0.056 & 0.054 & 0.052 & 0.053 \\ 
  & 400 & 0.000 & 0.000 & 0.000 & 0.056 & 0.056 & 0.058 & 0.059 & 0.056 & 0.052 & 0.056 \\ 
mix1 \\
  & 50 & 0.000 & 0.000 & 0.000 & 0.058 & 0.059 & 0.060 & 0.050 & 0.050 & 0.051 & 0.066 \\ 
  & 100 & 0.000 & 0.000 & 0.000 & 0.059 & 0.061 & 0.061 & 0.051 & 0.052 & 0.052 & 0.056 \\ 
  & 200 & 0.000 & 0.000 & 0.000 & 0.056 & 0.057 & 0.058 & 0.054 & 0.055 & 0.058 & 0.053 \\ 
  & 300 & 0.000 & 0.000 & 0.000 & 0.055 & 0.054 & 0.056 & 0.054 & 0.057 & 0.054 & 0.052 \\ 
  & 400 & 0.000 & 0.000 & 0.000 & 0.056 & 0.057 & 0.056 & 0.055 & 0.055 & 0.051 & 0.054 \\ 
mix2 \\
  & 50 & 0.000 & 0.000 & 0.000 & 0.062 & 0.063 & 0.067 & 0.056 & 0.057 & 0.057 & 0.067 \\ 
  & 100 & 0.000 & 0.000 & 0.000 & 0.054 & 0.056 & 0.057 & 0.052 & 0.055 & 0.054 & 0.055 \\ 
  & 200 & 0.000 & 0.000 & 0.000 & 0.054 & 0.055 & 0.055 & 0.053 & 0.055 & 0.055 & 0.054 \\ 
  & 300 & 0.000 & 0.000 & 0.000 & 0.051 & 0.053 & 0.052 & 0.053 & 0.055 & 0.051 & 0.052 \\ 
  & 400 & 0.000 & 0.000 & 0.000 & 0.050 & 0.050 & 0.051 & 0.057 & 0.055 & 0.052 & 0.050 \\ 
   \bottomrule
\end{tabular}
\end{table}

\begin{table}[ht]
\centering
\caption{Power rates by the nonparametric approach with various weight functions and by the logistic regression method with respect to the sample size and the parameter $\zeta$ for different sources of DIF.} 
\label{app:tab:simulation_PR}
\begin{tabular}{lr rrr rrr rrr r}
\toprule
  \multirow{3}{*}{\shortstack{\\[0.5ex]DIF\\[0.75ex] source}} & \multirow{3}{*}{$n$} & \multicolumn{9}{c}{Nonparametric} & \multirow{3}{*}{Logistic} \\ \cmidrule(lr){3-11}
   &  & \multicolumn{3}{c}{Optimal} & \multicolumn{3}{c}{Fixed} & \multicolumn{3}{c}{Bootstrap} &  \\ \cmidrule(lr){3-5} \cmidrule(lr){6-8} \cmidrule(lr){9-11}
 & & 0.260 & 0.292 & 0.320 & 0.260 & 0.292 & 0.320 & 0.260 & 0.292 & 0.320 & \\ 
\midrule
$a$ \\
   & 50 & 0.168 & 0.168 & 0.174 & 0.059 & 0.062 & 0.067 & 0.093 & 0.103 & 0.102 & 0.195 \\ 
   & 100 & 0.271 & 0.273 & 0.275 & 0.059 & 0.065 & 0.063 & 0.190 & 0.186 & 0.191 & 0.312 \\ 
   & 200 & 0.503 & 0.497 & 0.494 & 0.071 & 0.073 & 0.076 & 0.378 & 0.360 & 0.344 & 0.582 \\ 
   & 300 & 0.647 & 0.618 & 0.613 & 0.081 & 0.094 & 0.092 & 0.522 & 0.493 & 0.474 & 0.733 \\ 
   & 400 & 0.754 & 0.739 & 0.732 & 0.092 & 0.097 & 0.089 & 0.654 & 0.625 & 0.615 & 0.865 \\ 
$b$ \\
   & 50 & 0.308 & 0.305 & 0.293 & 0.309 & 0.317 & 0.314 & 0.219 & 0.210 & 0.198 & 0.270 \\ 
   & 100 & 0.486 & 0.485 & 0.478 & 0.500 & 0.510 & 0.505 & 0.382 & 0.357 & 0.341 & 0.446 \\ 
   & 200 & 0.775 & 0.769 & 0.763 & 0.796 & 0.801 & 0.789 & 0.659 & 0.631 & 0.602 & 0.742 \\ 
   & 300 & 0.904 & 0.893 & 0.889 & 0.923 & 0.922 & 0.911 & 0.829 & 0.809 & 0.779 & 0.906 \\ 
   & 400 & 0.961 & 0.961 & 0.959 & 0.968 & 0.968 & 0.966 & 0.935 & 0.925 & 0.900 & 0.967 \\ 
$c$ \\
   & 50 & 0.321 & 0.322 & 0.319 & 0.315 & 0.316 & 0.322 & 0.215 & 0.218 & 0.203 & 0.256 \\ 
   & 100 & 0.512 & 0.493 & 0.493 & 0.513 & 0.512 & 0.506 & 0.379 & 0.364 & 0.347 & 0.432 \\ 
   & 200 & 0.806 & 0.797 & 0.795 & 0.814 & 0.821 & 0.798 & 0.690 & 0.666 & 0.644 & 0.781 \\ 
   & 300 & 0.920 & 0.913 & 0.916 & 0.926 & 0.930 & 0.927 & 0.863 & 0.829 & 0.808 & 0.912 \\ 
   & 400 & 0.966 & 0.964 & 0.961 & 0.970 & 0.973 & 0.970 & 0.930 & 0.921 & 0.888 & 0.972 \\ 
$d$ \\
   & 50 & 0.306 & 0.306 & 0.293 & 0.299 & 0.296 & 0.290 & 0.230 & 0.232 & 0.229 & 0.289 \\ 
   & 100 & 0.465 & 0.462 & 0.466 & 0.455 & 0.454 & 0.450 & 0.371 & 0.369 & 0.358 & 0.463 \\ 
   & 200 & 0.760 & 0.751 & 0.746 & 0.752 & 0.765 & 0.741 & 0.660 & 0.636 & 0.615 & 0.766 \\ 
   & 300 & 0.866 & 0.864 & 0.858 & 0.873 & 0.875 & 0.867 & 0.793 & 0.765 & 0.741 & 0.898 \\ 
   & 400 & 0.957 & 0.949 & 0.942 & 0.959 & 0.946 & 0.939 & 0.918 & 0.892 & 0.872 & 0.970 \\ 
mix1 \\
   & 50 & 0.264 & 0.265 & 0.262 & 0.252 & 0.259 & 0.256 & 0.181 & 0.182 & 0.173 & 0.219 \\ 
   & 100 & 0.405 & 0.407 & 0.413 & 0.384 & 0.385 & 0.387 & 0.310 & 0.301 & 0.298 & 0.329 \\ 
   & 200 & 0.715 & 0.701 & 0.700 & 0.675 & 0.684 & 0.682 & 0.585 & 0.566 & 0.528 & 0.637 \\ 
   & 300 & 0.856 & 0.851 & 0.846 & 0.831 & 0.832 & 0.825 & 0.763 & 0.730 & 0.713 & 0.806 \\ 
   & 400 & 0.931 & 0.934 & 0.930 & 0.911 & 0.917 & 0.915 & 0.862 & 0.843 & 0.824 & 0.900 \\ 
mix2 \\
   & 50 & 0.115 & 0.121 & 0.130 & 0.089 & 0.094 & 0.090 & 0.075 & 0.081 & 0.081 & 0.123 \\ 
   & 100 & 0.201 & 0.202 & 0.194 & 0.137 & 0.142 & 0.142 & 0.163 & 0.147 & 0.150 & 0.174 \\ 
   & 200 & 0.328 & 0.317 & 0.310 & 0.192 & 0.200 & 0.197 & 0.227 & 0.229 & 0.219 & 0.251 \\ 
   & 300 & 0.453 & 0.434 & 0.427 & 0.285 & 0.286 & 0.290 & 0.346 & 0.336 & 0.319 & 0.400 \\ 
   & 400 & 0.565 & 0.552 & 0.548 & 0.334 & 0.331 & 0.326 & 0.464 & 0.436 & 0.401 & 0.503 \\ 
   \bottomrule
\end{tabular}
\end{table}

\begin{table}[ht]
\centering
\caption{MSE of the estimates of optimal weights with respect to the parameter $\zeta$, source of DIF, and sample size.} 
\label{app:tab:simulation_MSE}
\begin{tabular}{lrrrr}
  \toprule
  \multirow{2}{*}{\shortstack{\\[0.5ex]DIF\\[0.75ex] source}} & \multirow{2}{*}{$n$} & \multicolumn{3}{c}{Parameter $\zeta$} \\ \cmidrule(lr){3-5}
 &  & 0.260 & 0.292 & 0.320 \\ 
  \midrule
$a$ \\
   & 50 & 0.312 & 0.263 & 0.222 \\ 
   & 100 & 0.306 & 0.266 & 0.239 \\ 
   & 200 & 0.255 & 0.223 & 0.203 \\ 
   & 300 & 0.261 & 0.229 & 0.212 \\ 
   & 400 & 0.244 & 0.216 & 0.203 \\ 
$b$ \\
   & 50 & 0.026 & 0.033 & 0.042 \\ 
   & 100 & 0.008 & 0.009 & 0.011 \\ 
   & 200 & 0.005 & 0.006 & 0.005 \\ 
   & 300 & 0.005 & 0.005 & 0.005 \\ 
   & 400 & 0.004 & 0.004 & 0.004 \\ 
$c$ \\
   & 50 & 0.082 & 0.101 & 0.122 \\ 
   & 100 & 0.030 & 0.036 & 0.042 \\ 
   & 200 & 0.026 & 0.031 & 0.035 \\ 
   & 300 & 0.017 & 0.020 & 0.023 \\ 
   & 400 & 0.012 & 0.013 & 0.015 \\ 
$d$ \\
   & 50 & 0.060 & 0.066 & 0.074 \\ 
   & 100 & 0.021 & 0.022 & 0.024 \\ 
   & 200 & 0.007 & 0.007 & 0.008 \\ 
   & 300 & 0.006 & 0.007 & 0.005 \\ 
   & 400 & 0.007 & 0.006 & 0.005 \\ 
mix1 \\
   & 50 & 0.380 & 0.350 & 0.328 \\ 
   & 100 & 0.406 & 0.367 & 0.346 \\ 
   & 200 & 0.368 & 0.337 & 0.319 \\ 
   & 300 & 0.389 & 0.363 & 0.339 \\ 
   & 400 & 0.360 & 0.341 & 0.322 \\ 
mix2 \\
   & 50 & 1.006 & 0.956 & 0.915 \\ 
   & 100 & 0.928 & 0.882 & 0.832 \\ 
   & 200 & 0.905 & 0.865 & 0.836 \\ 
   & 300 & 0.849 & 0.804 & 0.779 \\ 
   & 400 & 0.822 & 0.779 & 0.754 \\ 
   \bottomrule
\end{tabular}
\end{table}

\newpage



\section{Figures}

\setcounter{figure}{0}
\renewcommand{\thefigure}{A\arabic{figure}}

\begin{figure}[hbt]
    \centering
    \includegraphics[width = 0.985\textwidth]{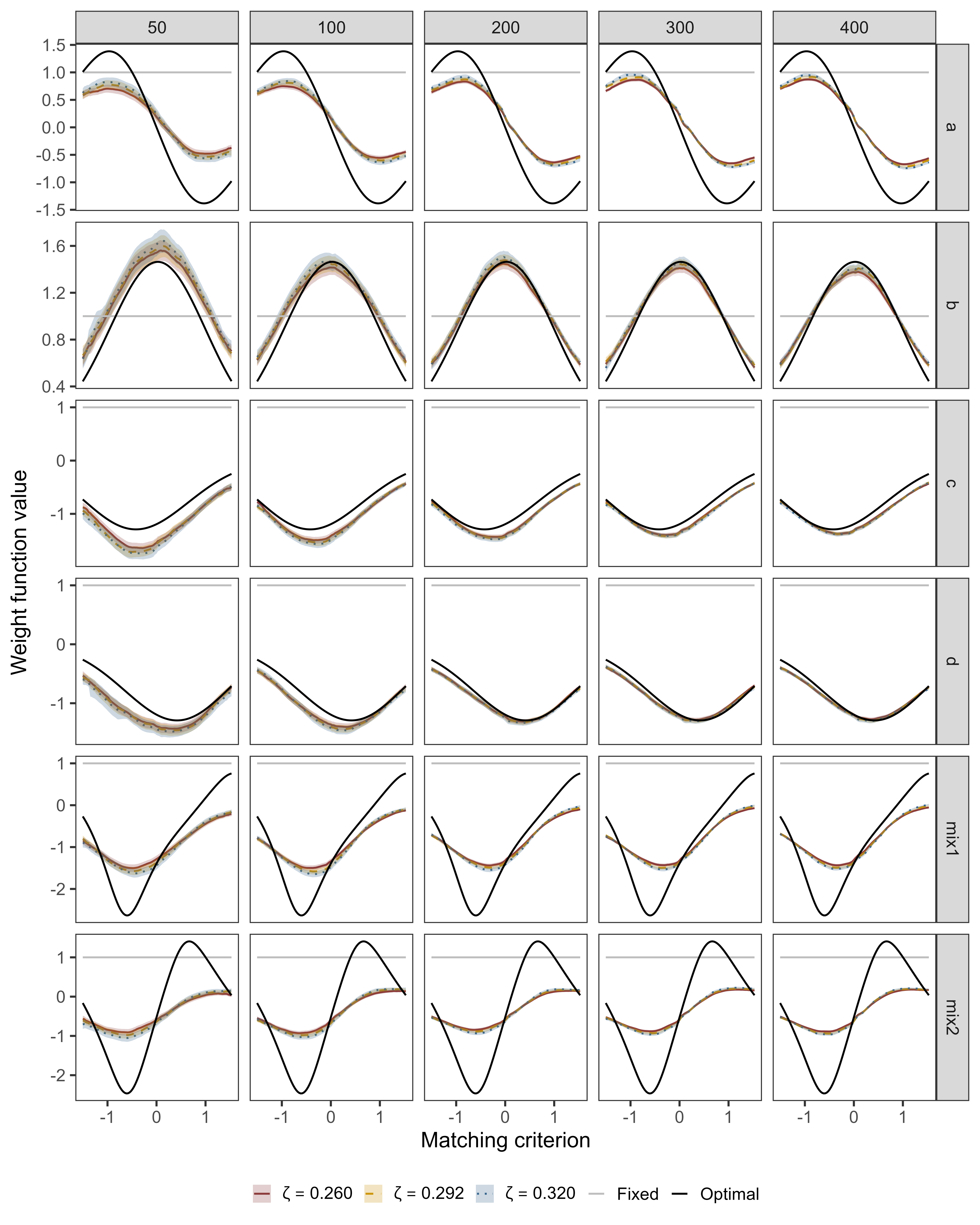}
    \caption{Estimates of optimal weights with confidence intervals with respect to the parameter $\zeta$, sample size, and source of DIF.}
    \label{app:fig:simulation_weights}
\end{figure}

\end{appendices}

\end{document}